\documentclass[letterpaper,usenatbib]{mn2e}
\usepackage{graphicx}
\usepackage{color}
\usepackage{lscape}
\usepackage{rotating}
\usepackage{longtable}
\usepackage{epstopdf}
\newsavebox{\tablebox}
\usepackage{hyperref}
\usepackage{times}
\usepackage{longtable,lscape}

\newcommand{\kms}{\ifmmode {\rm km\ s}^{-1} \else km s$^{-1}$\ \fi}
\newcommand{\ergs}{\ifmmode {\rm erg\ s}^{-1} \else erg s$^{-1}$\ \fi}

\newcommand{\feii}{Fe {\sc ii}\ }
\newcommand{\mgii}{Mg {\sc ii}\ }
\newcommand{\mmgii}{Mg {\sc ii}}
\newcommand{\civ}{C {\sc iv}\ }
\newcommand{\cciv}{C {\sc iv}}
\newcommand{\ciii}{C {\sc iii}\ }

\newcommand{\siiv}{Si {\sc iv}\ }

\newcommand{\lb}{\ifmmode L_{\rm Bol} \else $L_{\rm Bol}$\ \fi}
\newcommand{\ledd}{\ifmmode L_{\rm Edd} \else $L_{\rm Edd}$\ \fi}
\newcommand{\lx}{\ifmmode L_{\rm 2-10keV} \else  $L_{\rm 2-10keV}$\ \fi}
\newcommand{\hb}{\ifmmode H\beta \else H$\beta$\ \fi}
\newcommand{\ha}{\ifmmode H\alpha \else H$\alpha$\ \fi}
\newcommand{\heii}{He {\sc ii}\ }
\newcommand{\oiii}{[O {\sc iii}]\ }

\newcommand{\mbh}{\ifmmode M_{\rm BH}  \else $M_{\rm BH}$\ \fi}
\newcommand{\lv}{\ifmmode \lambda L_{\lambda}(5100\AA) \else $\lambda L_{\lambda}(5100\AA)$\ \fi}

\newcommand{\mdot}{\ifmmode \dot{m} \else \dot{m} \fi }
\newcommand{\llog}{\ifmmode {\rm log} \else {\rm log} \fi }
\newcommand\gtsima{$\; \buildrel > \over \sim \;$}
\newcommand\simgt{\lower.5ex\hbox{\gtsima}}

\newcommand\lbol{{$L_{\rm Bol}$}}
\newcommand{\leddR}{\ifmmode L_{\rm Bol}/L_{\rm Edd} \else $L_{\rm Bol}/L_{\rm Edd}$\ \fi}
\def\cm2{{cm$^{-2}$}}

\newcommand{\aox}{$\alpha_{\rm ox}$}
\begin{document}
\title[The \civ Baldwin effect]{The underlying driver for the \civ Baldwin effect in QSOs with $0<z<5$}
\author[X. Ge et al.]{Xue Ge, Wei-Hao Bian \thanks{E-mail: whbian@njnu.edu.cn}, Xiao-Lei Jiang, Wen-Shuai Liu and Xiao-Feng Wang \\
Department of Physics and Institute of Theoretical Physics, Nanjing
Normal University, Nanjing 210023, China\\} \maketitle

\begin{abstract}
Broad emission lines is a prominent property of type I quasi-stellar objects (QSOs). The origin of the Baldwin effect for \civ $\lambda1549~$\AA\ broad emission lines, i.e., the luminosity dependence of the \civ equivalent width (EW), is not clearly established. Using a sample of 87 low-$z$ Palomar-Green (PG) QSOs and 126 high-$z$ QSOs across the widest possible ranges of redshift ($0<z<5$), we consistently calculate \hb-based single-epoch supermassive black hole (SMBH) mass and the Eddington ratio to investigate the underlying driver of the \civ Baldwin effect. An empirical formula to estimate the host fraction in the continuum luminosity at 5100 \AA\ is presented and used in \hb-based \mbh calculation for low-$z$ PG QSOs. It is found that, for low-$z$ PG QSOs, the Eddington ratio has strong correlations with PC1 and PC2 from the principal component analysis, and \civ EW has a strong correlation with the optical \feii strength or PC1. Expanding the luminosity range with high-$z$ QSOs, it is found that \civ Baldwin effect exists in our QSOs sample. Using \hb-based single-epoch SMBH mass for our QSOs sample, it is found that \civ EW has a strong correlation with the Eddington ratio, which is stronger than that with the SMBH mass. It implies that the Eddington ratio seems to be a better underlying parameter than the SMBH mass to drive the \civ Baldwin effect.
\end{abstract}

\begin{keywords}
black hole physics --- galaxies:active --- quasars:emission lines
\end{keywords}

\section{INTRODUCTION}
Broad emission lines is a prominent property of type I active galactic nuclei (AGNs) and quasi-stellar objects (QSOs). It is accepted that these broad emission lines are produced by photoionization in broad-line regions (BLRs) gas, where the accretion disc surround the central supermassive black hole (SMBH) provides the ionizing photos. \cite{Baldwin1977} discovered an anti-correlation between the equivalent width (EW) of the \civ $\lambda1549~$\AA\ emission line and the continuum luminosity in the QSO rest frame, \citep[i.e. the Baldwin effect, see the review by][]{Shields2007}. Over the past 20 yr, this effect was investigated with different
QSOs samples, including other permitted/prohibited emission lines, such as Ly$\alpha$, \ciii, \siiv, \mgii, \oiii, Fe K$\alpha$ \citep[e.g.,][]{Green2001, Dietrich2002, Shang2003, Baskin2004, Xu2008, Wu2009, Richards2011, Bian2012, Shen2014}.

It is believed that the Baldwin effect exists for many UV/optical emission lines. However, its origin is not clearly established. One promising interpretation is the softening of the spectral energy distribution (SED) for increasing luminosity, which lowers the ion populations that having high ionization potentials \citep[e.g.,][]{Netzer1992, Dietrich2002}. Non-isotropic continuum emission and the intrinsic Baldwin effect would provide some of the observed scatter in the Baldwin effect \citep[e.g.,][]{Baskin2004}. The underlying physical parameters for the Baldwin effect are investigated for many years, such as the eigenvector 1 of \cite{Boroson1992}, the Eddington ratio \leddR (i.e. the ratio of the bolometric luminosity to the Eddington luminosity), the SMBH mass \mbh \citep[e.g.,][]{Boroson1992, Wills1999,Boroson2002, Shang2003, Bachev2004,Baskin2004, Xu2008, Bian2012, Shemmer2015}. For a optical-selected sample of Palomar-Green (PG) QSOs, \cite{Baskin2004} found a strong correlation of the \civ EW with \leddR, stronger than that with the continuum luminosity, and suggested that the \leddR is the primary physical parameter which drives the \civ Baldwin effect \citep[]{Shemmer2015}. Using a larger sample of QSOs with $1.5<z<5$ from Sloan Digital Sky Survey (SDSS), \cite{Xu2008} found the \civ EW has a stronger correlation with \cciv-based \mbh than the \leddR. Using \cciv-based \mbh by \cite{Shen2011} for high-$z$ SDSS QSOs, \cite{Bian2012} also found that there is a correlation between the \civ EW and the \civ-based \mbh. However, with \mmgii-based \mbh, \cite{Bian2012} found SDSS QSOs in $1.5<z<1.9$ follow the relation found by \cite{Baskin2004}, suggesting the bias in \cciv-based \mbh.

Due to the larger consumption of telescope time in reverberation mapping (RM), there is about 60 AGN/QSOs with reliable BLRs sizes from RM \citep[]{Kaspi2000,Peterson2004,Du2016, Shen2016}. From the RM BLRs sizes, there is an empirical R-L relation \citep[]{Kaspi2000,Bentz2013, Du2015}. With this empirical R-L relation, the BLRs size can be derived from the continuum luminosity at 5100 \AA. The single-epoch SMBH mass can be calculated from the broad emission lines, such as \hb, \ha, \mgii, \civ \citep[e.g.,][]{Laor1998, Bian2004, Greene2005, Vestergaard2006, Jun2015}. By the host correction in the continuum luminosity at 5100 \AA\ by Hubble Space Telescope (HST) images for different width slits, \cite{Bentz2013} found that the slope in R-L relation changes from 0.7 to 0.533. It is suggested that \civ-based \mbh is biased to the \hb-based \mbh \citep[e.g.,][]{Rafiee2011, Shen2012, Bian2012}. For high-$z$ QSOs, the \hb emission line is shifted to the infrared (IR) band. The IR spectroscopy observation for high-$z$ QSOs is needed to calculate SMBH \mbh using the same \hb emission line as that for low-$z$ QSOs.

In order to investigate the \civ EW relation with the SMBH accretion, we compile a sample of 87 low-$z$ PG QSOs and 126 high-$z$ QSOs across the widest possible ranges of redshift ($0<z<5$) with available spectral information for \hb $\lambda4861$ \AA\ and \civ $\lambda1549$ \AA\ emission lines. Our adopted sample is described in \S 2, the results and the analysis are given in \S 3, and the conclusions are presented in \S 4. All of the cosmological calculations in this paper assume $H_{0}=70 \rm {~km ~s^ {-1}~Mpc^{-1}}$, $\Omega_{M}=0.3$, and $\Omega_{\Lambda} = 0.7$.

\section{Sample}

\begin{figure}
\begin{center}
\includegraphics[height=5cm]{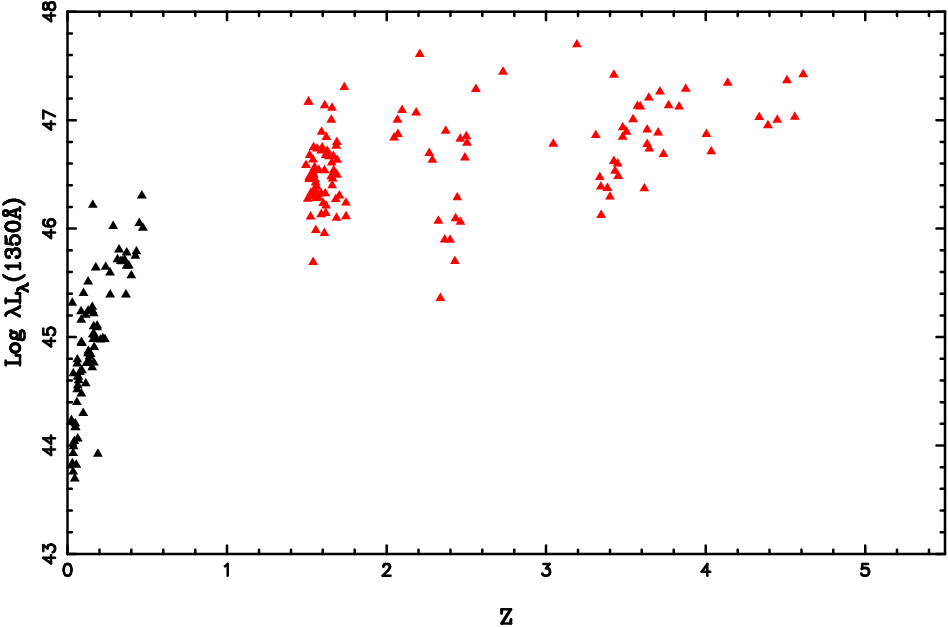}
\caption{The UV continuous luminosity at 1350\AA\ (in units of erg/s) versus redshift for our low-$z$ and high-$z$ sample. The black triangles denote 87 low-$z$ PG QSOs, and the red triangles denote 126 high-$z$ QSOs.}
\label{fig1}
\end{center}
\end{figure}

\subsection{low-$z$ sample}
The 87 PG QSOs ($0<z<0.5$) are optically selected by a limiting $B$-band magnitude of 16.16, blue $U-B$ colour ($<-0.44$), and dominant starlike appearance, showing broad emission lines classified as type 1 QSOs  \citep{Schmidt1983, Boroson1992}. The low-$z$ PG QSOs is representative of bright optically selected QSOs \citep{Jester2005}. It is the most thoroughly explored sample of AGN/QSOs, with a lot of high quality data at most wave bands \citep[e.g.,][]{Boroson1992, Brandt2000, Baskin2004, Bian2014, Shi2014}.
\cite{Boroson1992} observed all these 87 PG QSOs with the KPNO 2.1 m telescope and the Gold Spectrograph. The spectra of these objects were made through a $1.5$ arcsec slit with a spectral resolution of about 6.5 \AA, covering the range 4300-5700 \AA\ in the rest frame. In this paper, the \hb width at half-maximum (FWHM) is adopted from \cite{Boroson1992} used in the \hb-based single epoch SMBH mass calculation \citep[see also][]{Boroson2002, Vestergaard2006}. In Table 1, the spectral-resolution-corrected \hb FWHM adopted from \cite{Vestergaard2006} is listed in Col. (9). For UV spectra of these 87 PG QSOs, \cite{Baskin2004} obtained archived UV spectra for these 85 PG QSOs, 47 from the HST and 38 from the International Ultraviolet Explorer (IUE). For three of them (PG 0934+013, PG 1004+130, PG 1448+273), their UV archival spectra did not have a sufficient S/N to measure the \civ EW, and PG 1700+518 is a broad-absorption line (BAL) QSOs. In addition, there are five BAL QSOs and 16 radio-loud (RL) QSOs. The continuum luminosity at 1350 \AA\ / 5100 \AA\ is calculated from the (SED) presented by \cite{Neugebauer1987} \citep[]{Baskin2004,Vestergaard2006}. Table 1 lists the information of 87 PG QSOs. Fig.~\ref{fig1} shows $\lambda L_{\lambda}(1350\AA)$ versus $z$, where black triangles denote 87 PG QSOs.


\subsection{High-$z$ sample}
In this paper, the \hb emission line is needed to calculate the single-epoch \mbh for all QSOs and to investigate the underlying parameter for the \civ Baldwin effect. For high-$z$ QSOs with available \civ $\lambda 1459$ \AA\ observed by ground telescope ($z>1.5$), the \hb emission line is shifted to the IR band.

\cite{Shen2012} presented 60 intermediate-redshift QSOs ($z\sim 1.5-2.2$) selected from SDSS DR7. Their near-IR spectrum are observed with TripleSpec \citep{Wilson2004} ($0.95-2.46~\mu m$) on the ARC 3.5 m telescope, and with the Folded-port InfraRed Echellette \citep[]{Simcoe2010} ($0.8-2.5~\mu m$) on the 6.5\,m Magellan-Baade telescope. With TripleSpec, the total exposure is typically $1-1.5$\,hr with slits widths of both 1.1\arcsec\ and 1.5\arcsec, and the resulting spectral resolution of $R\sim 2500-3500$.  With FIRE, typical total exposure times were 45\,min with the slit width of 0.6\arcsec. The spectral resolution is $R\sim 6000$ (50\,${\rm km\,s^{-1}}$). They performed standard ABBA dither patterns to aid sky subtraction and observed a nearby A0V star as flux and telluric standard. \cite{Jun2015} presented near-Infrared grism spectra for 155 QSOs ($3.3<z<6.4$) from the $AKARI$ space telescope. It is composed of optically luminous and spectroscopically confirmed type 1 QSOs at $z>3$, mostly out of SDSS DR5. The spectrum coverage is $2.5-5.0~\mu m$. The spectral resolution is $R=120$ at $3.6 \mu m$, corresponding to a velocity resolution of 2500 \kms. \cite{Jun2015} gave 43 high-$z$ QSOs with \ha information. From above two literatures and others \citep[]{Shemmer2004, Netzer2007, Dietrich2009, Assef2011, Ho2012, Shen2012, Jun2015, Shemmer2015}, we assembled a sample of 182 high-$z$ QSOs with \hb/\ha data.

In order to obtain the UV \civ data for these high-$z$ QSOs, we search their SDSS spectral data. The SDSS used a 2.5-m wide-field telescope at Apache Point Observatory near Sacramento Peak in Southern New Mexico to conduct
an imaging and spectroscopic survey. \cite{Shen2011} presented a compilation of properties of the 105,783 QSOs in the SDSS DR7 catalogue, where they used multi-Gaussians to fit the main emission lines, such as \civ, \mgii, \hb, \ha depending on the redshift. We search these 182 high-$z$ QSOs in the SDSS DR7 QSOs sample by \cite{Shen2011}, and obtain 126 high-$z$ QSOs with \civ fitting data and the luminosity at 1350 \AA. \cite{Shen2011} suggested that measurements of \civ EW for most objects are unbiased to within 20\% down to S/N $\sim 3$. For our SDSS high-z sample, their S/N are larger than 3. There are 7 BAL QSOs, 12 RL QSOs, 11 weak line QSOs ($\rm EW(C IV) < 10\AA$). For these 126 high-$z$ QSOs, we respectively select 10, 15, 7, 1, 60, 7, and 26 high-$z$ QSOs from the literatures of \cite{Shemmer2004, Netzer2007, Assef2011, Ho2012, Shen2012, Shemmer2015, Jun2015}. For 26 high-$z$ QSOs in \cite{Jun2015}, we obtain the \lv and \ha FWHM from table 7 in \cite{Jun2015}. We convert the \ha FWHM to \hb FWHM by the formula \citep[]{Jun2015},
\begin{eqnarray}
\rm log~\frac{FWHM(\hb)}{1000~\kms}) = (1.061\pm 0.013)log~\frac{FWHM(\ha)}{1000~\kms}\,\nonumber\\+(0.055\pm 0.008).
\label{eq1}
\end{eqnarray}
The FWHM of \ha or \hb is listed in Col. (7) in Table 2. These 126 high-$z$ QSOs are used as our final high-$z$ QSOs sample with available \hb and \civ data at the same time. It is larger than the high-$z$ sample of 36 QSOs ($z<3.5$) by \cite{Shemmer2015}. Our sample covers the redshift of $0<z<5$ and the $log \lambda L_{\lambda}(1350\AA)$ of $43.6-47.7$. Table 2 lists the properties of high-z QSOs. In Fig.~\ref{fig1}, the red triangles denote these 126 high-$z$ QSOs.

\section{Result and Discussion}
\subsection{The host contribution in the continuum luminosity at 5100 \AA}

\begin{figure}
\begin{center}
\includegraphics[height=5cm]{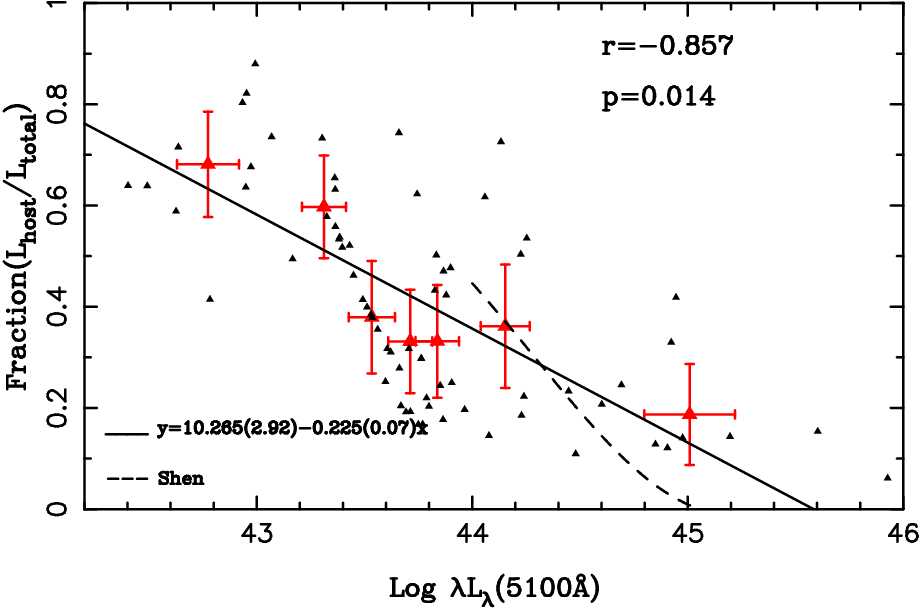}
\caption{The host fraction in the total continuum luminosity at 5100 \AA\ versus the total continuum luminosity at 5100 \AA. The black triangles are the observation data from \citep[]{Bentz2013}, the red triangle denotes the value of each bin. The black solid line is the best fit for bin data. The dotted line is the formula from \citep[]{Shen2011}. For red bin data, the Spearman correlation coefficient and the probability of the null hypothesis are shown in the panel.}
\label{fig2}
\end{center}
\end{figure}

With the empirical R-L relation, the BLRs size can be derived from the nuclei continuum luminosity. Based on the stacked SDSS spectra, \cite{Shen2011} gave an empirical formula to give the ratio of host to nuclei QSOs luminosity for the total QSOs continuum luminosity of $log (\lv/erg~s^{-1}) = 44.1-45.5$. When total QSOs luminosity are larger than $10^{45} erg/s$, the host contribution is negligible. For high-$z$ QSOs in our sample, we found they are all $log (\lv/erg~s^{-1})> 45$, and the host contamination is generally negligible. For low-$z$ PG QSOs, we need to correct the host contribution to derive the BLRs sizes because of their lower luminosity at 5100\AA.

Using GALFIT \citep[]{Peng2002} to model HST host-galaxy images of 41 RM AGN \cite{Bentz2013} did the host correction for 71 observation data at 5100 \AA. For the host fraction in the total continuum luminosity from \citep[]{Bentz2013}, we bin the observation data into seven bins according the total continuum luminosity at 5100 \AA. The value of each bin is the average of total continuum luminosity and the host fraction, and the error of each bin value is the standard deviation. For the bin data, the Spearman correlation test gives the Spearman correlation coefficient $r_{s}=0.86$ and the probability of the null hypothesis $P_{\rm null}=1.4\times 10^{-2}$. The relationship is derived by using algorithms FITEXY \citep[]{Press1992}.
\begin{eqnarray}
f=(10.265\pm 2.92)-(0.225\pm 0.07)~log \lv^{total}
\label{eq2}
\end{eqnarray}
where $f$ is the host fraction in the total continuum luminosity at 5100 \AA, $\lv^{total}$ is the total continuum luminosity at 5100 \AA. In Fig. ~\ref{fig2}, we also show the host fraction formula of \cite{Shen2011} for SDSS fibre, which is consistent with our formula, considering the decreasing tendency with large \lv. The host fraction decreases from about 80\% at $\lv=10^{42} erg/s$ to about 10\% at $\lv=10^{45} erg/s$.

\begin{figure}
\begin{center}
\includegraphics[height=5cm]{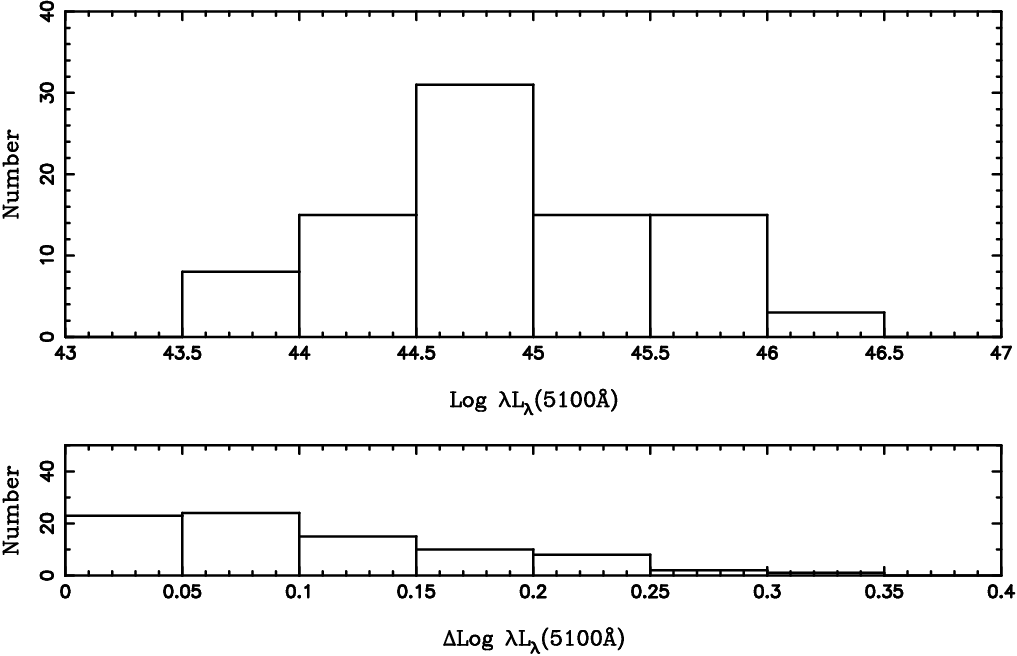}
\caption{Top: distribution of total continuum luminosity at 5100 \AA. Bottom: distribution of the difference between the total continuum luminosity and the nuclei continuum luminosity at 5100 \AA\ in logarithm, i.e., $log(1/(1-f))$. }
\label{fig3}
\end{center}
\end{figure}

For 87 low-$z$ PG QSOs, there are 16 RM objects, where the host fractions are adopted from \cite{Bentz2013}. For other PG QSOs with the total continuum luminosity \lv below $10^{45.5} erg/s$, we use our formula to correct the host contribution. The host-corrected \lv is listed in Col. (8) in Table 1. Fig. ~\ref{fig3} shows the distribution of total luminosity at 5100 \AA\ (top panel) for 87 PG QSOs and the distribution of the difference between the total continuum luminosity and the nuclei continuum luminosity at 5100 \AA\ in logarithm, i.e., $log(1/(1-f))$ (bottom panel). Considering the range of log \lv of $43.6-46.2$, the host fraction $f$ in the total luminosity at 5100\AA\ is estimated to be less than about 50\% from our formula of Eq. ~\ref{eq2}. $log(1/(1-f))=0.3$ for $f=50\%$ (bottom panel in Fig. ~\ref{fig3}).


\subsection{\mbh and \leddR }

\begin{figure}
\begin{center}
\includegraphics[height=5cm]{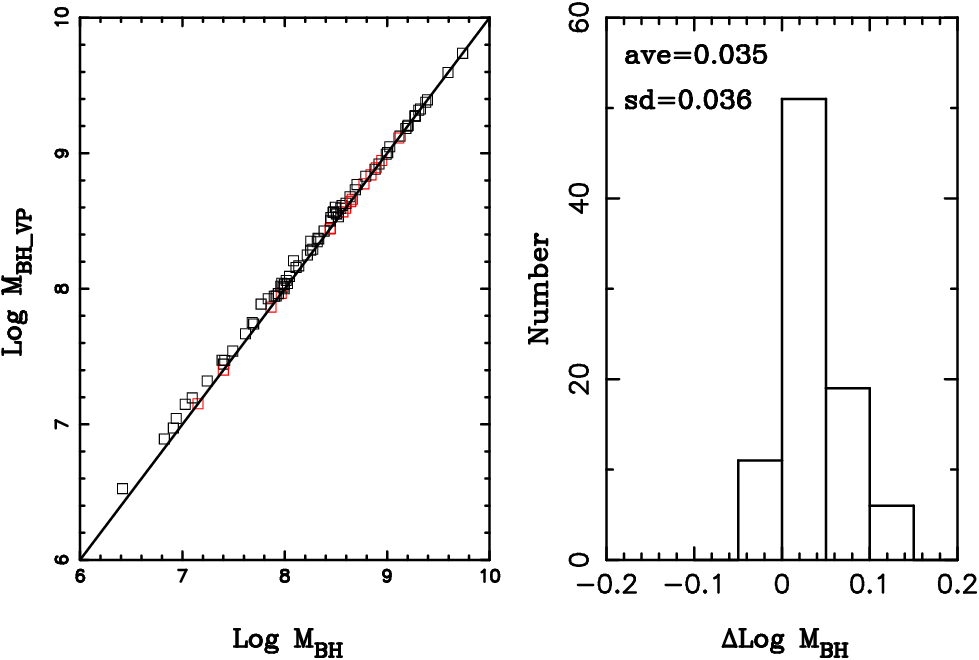}
\caption{Left: \mbh given in \citep{Vestergaard2006} versus our calculated \mbh considering host contribution for 87 low-$z$ PG QSOs. The red square denotes 16 RM objects. The black line is 1:1. Right: distribution of the \mbh difference.}
\label{fig4}
\end{center}
\end{figure}

\begin{figure}
\begin{center}
\includegraphics[height=5cm]{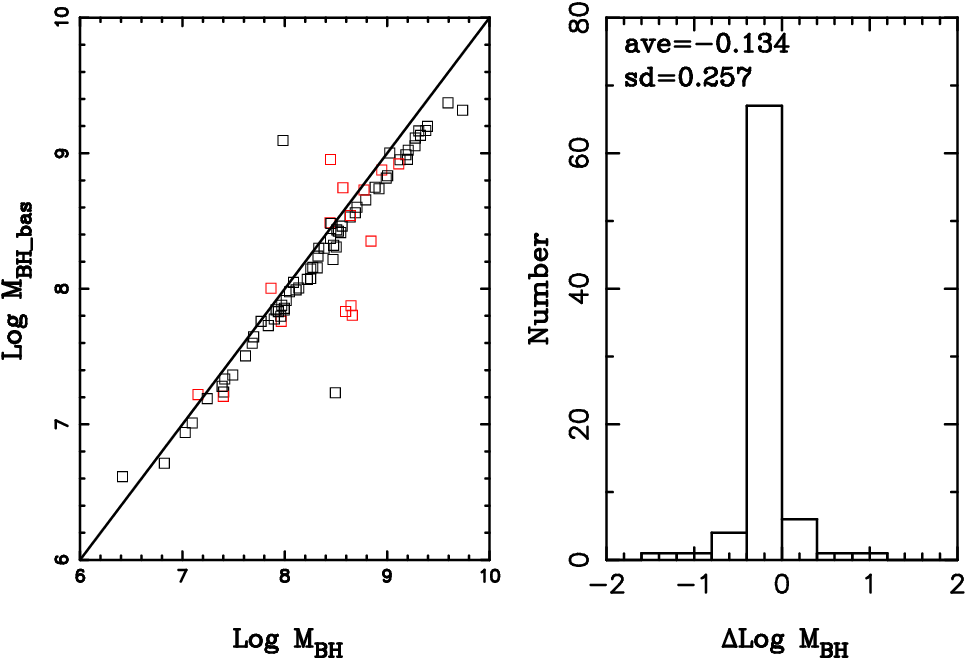}
\caption{Left: \mbh given in \citep{Baskin2004} versus our calculated \mbh considering host contribution for 87 low-$z$ PG QSOs. The red square denotes 16 RM objects. The black line is 1:1. Right: distribution of the \mbh difference.}
\label{fig5}
\end{center}
\end{figure}

Using 32 RM SMBHs masses from \cite{Peterson2004}, \cite{Vestergaard2006} gave a formula to calculate the SMBH mass from the AGN/QSOs single-epoch spectrum, where \hb FWHM and the continuum luminosity at 5100 \AA\ are measured.
\begin{eqnarray}
{log \mbh(\hb)}=log \{(\frac{\rm
FWHM(\hb)}{1000~\kms})^2 \times (\frac{\rm \lv}{10^{44}~\ergs})^{0.5}\}\,\nonumber\\+(6.91\pm0.02)
\label{eq3}
\end{eqnarray}
With above formula, we calculate the black hole mass using FWHM(\hb) and host-corrected continuum luminosity at 5100 \AA\ for 71 non-RM low-$z$ PG QSOs and 126 high-$z$ QSOs (see Tables 1 and 2). The host-corrected \mbh is listed in Col. (10) in Table 1 for low-z PG QSOs. We also list the \mbh given by \cite{Vestergaard2006} in Col. (11) and that by \cite{Baskin2004} in Col. (12) in Table 1. Fig. ~\ref{fig4} shows \mbh given in \cite{Vestergaard2006} (without host correction, same formula) versus our calculated \mbh considering host correction for 87 low-$z$ PG QSOs (left-hand panel) and the distribution of their \mbh difference (right-band panel). The red squares denote RM objects. The mean value of their difference is 0.035 with the standard deviation of 0.036. Considering that $\mbh \propto L^{0.5}$, the log\lv correction of 0.1 dex would lead to a decrease of 0.05 dex, which is consistent with the result in Fig. ~\ref{fig4}. With a different formula by \cite{Laor1998}, \cite{Baskin2004} also calculated the \mbh for 87 PG QSOs. Fig. ~\ref{fig5} shows our calculated \mbh versus that in \cite{Baskin2004} (left-hand panel) and the distribution of their difference (right-hand panel). The mean value of their difference is 0.134 dex with the standard deviation of 0.257. Considering intrinsic scatter in \mbh calculation of about 0.3 dex \citep[e.g.,][]{Shen2011}, the host correction in \mbh calculation is negligible for low-$z$ PG QSOs sample, although it is important for low-luminosity AGN/QSOs. For high-z QSos, the \mbh is also listed in Col. (9) in Table 2.

We use the nuclei continuum luminosity at 5100\AA\ to calculate the bolometric luminosity by a bolometric correction of 9.26 \citep{Richards2006}, and then calculate the Eddington ratio \leddR for our low-$z$ QSOs. For high-$z$ QSOs, \lv is so large that the host contribution can be neglected, and \leddR are calculated as that for low-$z$ PG QSOs. \leddR for high-$z$ QSOs is listed in Col. (10) in Table 2.

\subsection{The relation between the SMBH accretion and PC1/PC2 from PCA}
\begin{figure}
\begin{center}
\includegraphics[height=5cm]{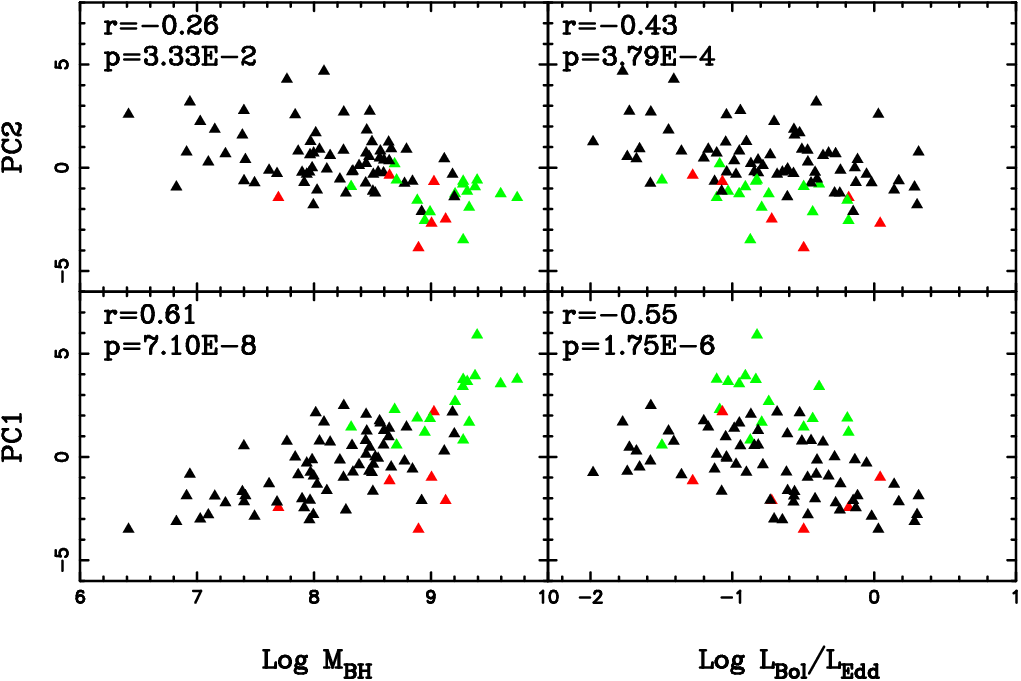}
\caption{Left: the relationship between \mbh and PC1, PC2. Right: the relationship between Eddington ratio \leddR and PC1, PC2. Green triangles denote RL QSOs. Red triangles denote BAL QSOs. The Spearman correlation coefficient and the probability of the null hypothesis  is shown in the left in each panel.}
\label{fig6}
\end{center}
\end{figure}

\begin{figure}
\begin{center}
\includegraphics[height=5cm]{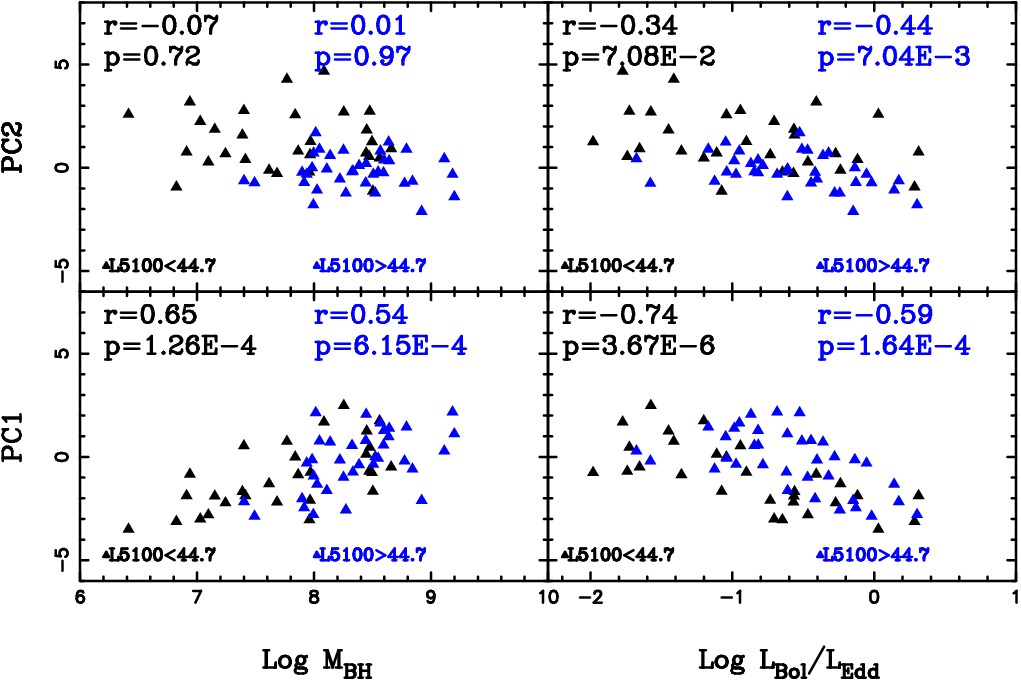}
\caption{The same as Fig. ~\ref{fig6}~ but for two subsamples dividing by luminosity $log\lv = 44.7$ (excluding RL QSOs and BAL QSOs). The back triangles denote PG QSOs with smaller \lv and blue triangles denote PG QSOs with larger \lv. The Spearman correlation coefficient and the probability of the null hypothesis is shown in each panel.}
\label{fig7}
\end{center}
\end{figure}

\begin{figure}
\begin{center}
\includegraphics[height=5cm]{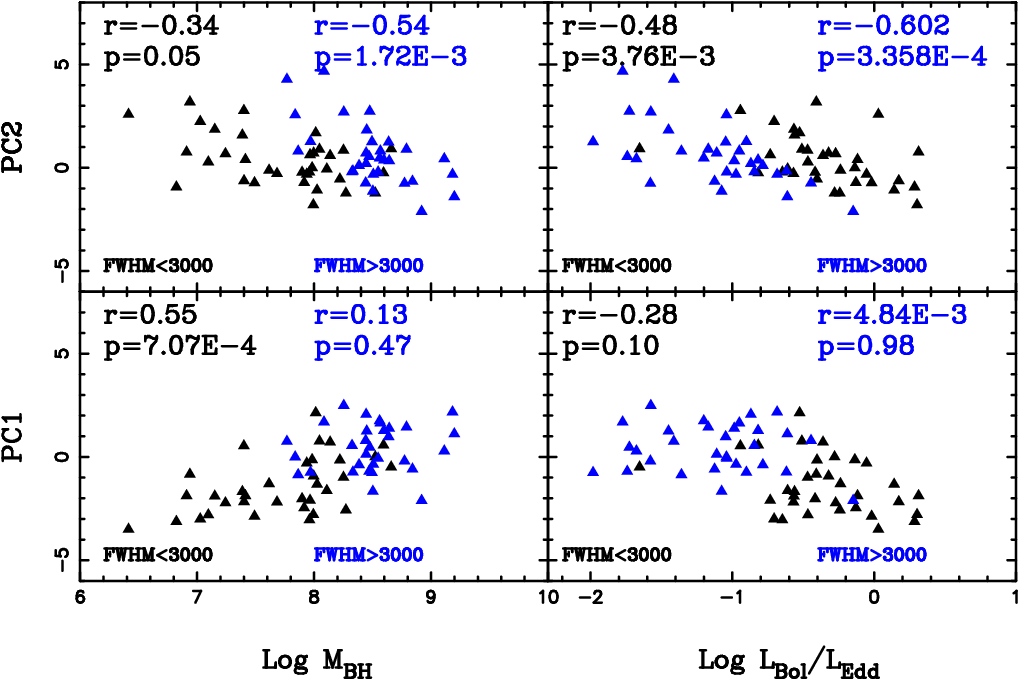}
\caption{Same as Fig. ~\ref{fig6}~ but for two subsamples dividing by $FWHM(\hb) = 3000 \kms$ (excluding RL QSOs and BAL QSOs). The back triangles denote PG QSOs with smaller \hb FWHM and blue triangles denote PG QSOs with larger \hb FWHM. The Spearman correlation coefficient and the probability of the null hypothesis  is shown in each panel.}
\label{fig8}
\end{center}
\end{figure}

With the optical spectral information and additional information from other bands for 87 low-$z$ PG QSOs, \cite{Boroson1992} presented principal component analysis (PCA) and found that the variance in the optical emission lines and the continuum (radio through X-ray) was mostly contained in two sets of correlations, eigenvectors of the correlation matrix. Principal component 1 (PC1) links the strength of optical \feii emission, \oiii emission, and \hb line asymmetry. Principal component 2 (PC2) involves optical luminosity and the strength of \heii 4686 and \aox. With the single-epoch SMBH mass derived from \hb \citep{Kaspi2000}, \cite{Boroson2002} found that PC1 is mainly correlation with \leddR and PC2 has a strong correlations with \mbh and \leddR. He suggested that PC1 is driven predominantly by the Eddington ratio, and PC2 is driven by the accretion rate. The coefficient listed in table 1 in \cite{Boroson2002} is used to calculate the PC1/PC2 values. PC1/PC2 values are listed in Col. (15-16) in Table 1. Fig. ~\ref{fig6} shows PC1/PC2 versus our calculated \mbh and \leddR for 87 low-$z$ PG QSOs. We find that PC1 has strong correlations with \mbh and \leddR, r=0.66, -0.46, respectively. PC2 has a strong correlation with \mbh and a weak correlation with \leddR, r=-0.53, -0.34, respectively. These results are consistent with \cite{Boroson2002}. In Fig. ~\ref{fig6}, we also show 16 RL QSOs (green triangle) and 5 BAL QSOs (red triangle). Considering the speciality RL QSOs and BAL QSOs, we exclude them and find that PC1 still has strong correlation with \mbh and \leddR, r=0.61, -0.55, respectively. PC2 has a weak correlation with \mbh and a strong correlation with \leddR, r=-0.26, -0.43, respectively. It is different to the results by \cite{Boroson2002}, where PC2 has a very weak correlation with \leddR and a strong correlation with \mbh. With our calculated \mbh and \leddR and excluding RL QSO ang BAL QSO, we find that PC2 has a strong correlation with \leddR and a weak correlation with \mbh. We think it is due to narrower PC1/PC2 parameter space (see Fig. ~\ref{fig6}).

In Fig. ~\ref{fig7}, we divide low-$z$ PG QSOs into two parts at $log\lv=44.7$. For low-luminosity part, PC1 has strong correlations with \mbh and \leddR. PC2 has weak correlations with \mbh and \leddR. For high-luminosity part, PC1 still has strong correlations with \mbh and \leddR. PC2 has a weak correlation with \mbh and a strong correlation with \leddR. We find that there is no correlation between \mbh and PC2 no matter at low luminosity or high luminosity. It is suggested that there exists the BLRs orientation effect to derive the BLR velocity from FWHM \citep[]{Shen2014}. In Fig. ~\ref{fig8}, we divide low-$z$ PG QSOs into two parts at $\rm FWHM(\hb)=3000~\kms$. For low-FWHM part, PC1 has a strong correlation with \mbh and a weak correlation with \leddR. PC2 has weak correlations with \mbh and \leddR. For high-FWHM part, PC1 still has very weak correlations with \mbh and \leddR. PC2 has strong correlations with \mbh and \leddR. The dividing with FWHM affects seriously the correlations between PC1 and \mbh and \leddR. Because PC1 has a strong correlation with \hb FWHM, dividing PG QSOs based on \hb FWHM would lead to narrower PC1 parameter space, which would lead to weaken PC1 correlation with \mbh and \leddR. PC2 also has a strong correlation with continuum luminosity, dividing PG QSOs based on the luminosity would lead to narrower PC2 parameter space, which would lead to weaken PC2 correlation with \mbh and \leddR. Therefore, considering the PG QSOs subsample with different radio loudness, continuum luminosity, and \hb FWHM, the narameter range limitation will change the strongness of the relations between PC1/PC2 and \mbh, \leddR. Large sample with larger parameter coverage is needed for this kind of work in the future.

\begin{figure}
\begin{center}
\includegraphics[height=5cm]{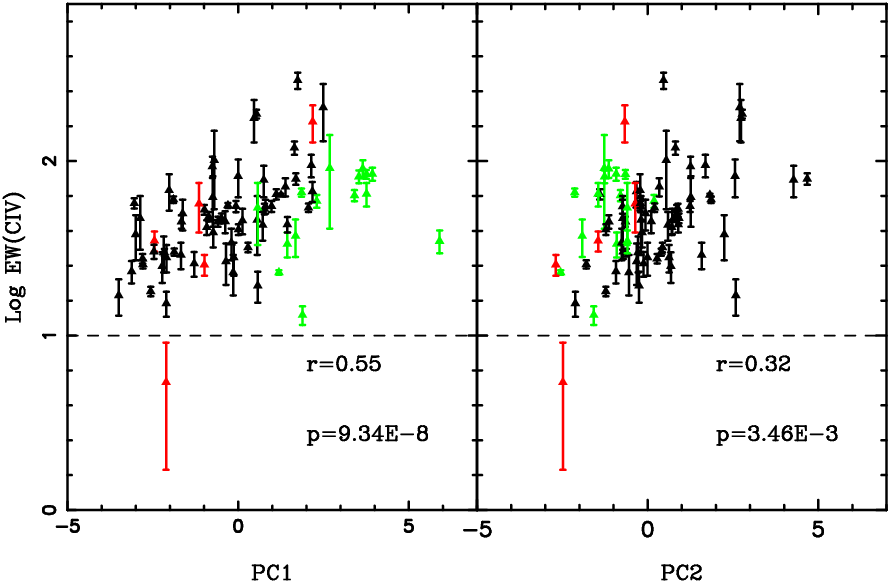}
\caption{Left: the relationship between \civ EW and PC1. Right: the relationship between \civ EW and PC2.
The marks are the same as Fig.~\ref{fig6}}
\label{fig9}
\end{center}
\end{figure}

Since PC1/PC2 has a correlation with \mbh or \leddR, in Fig. ~\ref{fig9}, we show the relationship between
\civ EW and PC1/PC2. The \civ EW has a strong correlation with PC1 and the correlation with PC2 is a little weaker. Excluding RL QSOs and BAL QSOs, these correlations become stronger.
From PCA shown in \cite{Boroson2002}, PC1 has a strong correlation with the \feii strength, $R_{\rm Fe}$ (the ratio of optical \feii and \hb EW). We find \civ EW also has a strong correlation with $R_{\rm Fe}$ with $r=-0.59$.

\subsection{Physical driver of the Baldwin effect of \civ}

\begin{figure}
\begin{center}
\includegraphics[height=5cm]{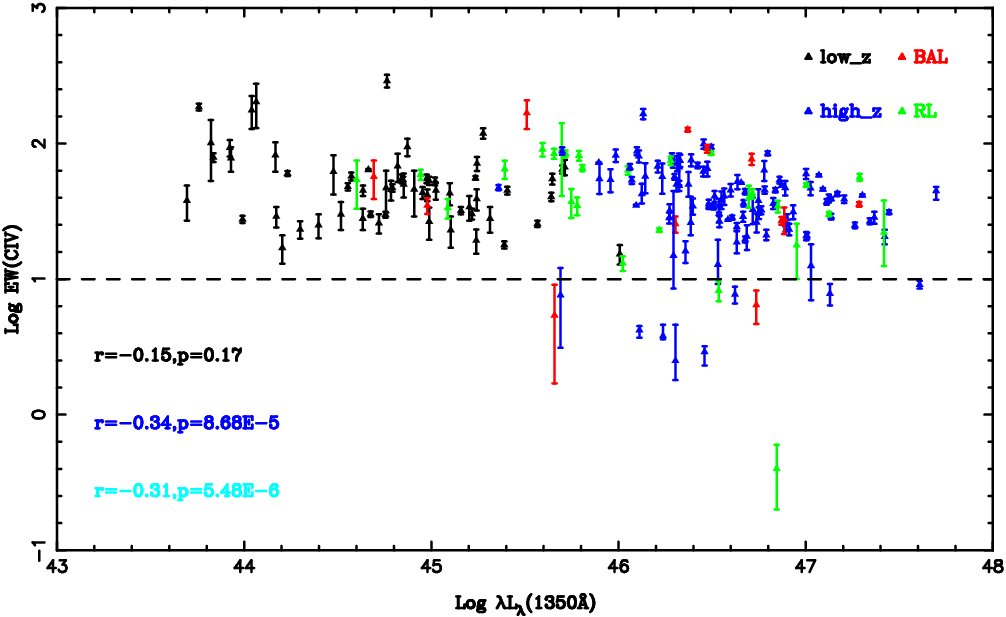}
\caption{The \civ Baldwin effect, i.e., the \civ EW versus the UV continuous luminosity at 1350\AA\ for our sample of low-$z$ and high-$z$ QSOs. Black triangles denote 81 PG QSOs with low redshift ($z<0.5$), and blue triangles denote SDSS QSOs ($1.5< z <5$). Red triangles denote BAL QSOs and green triangles denote RL QSOs.}
\label{fig10}
\end{center}
\end{figure}

\begin{figure}
\begin{center}
\includegraphics[height=5cm]{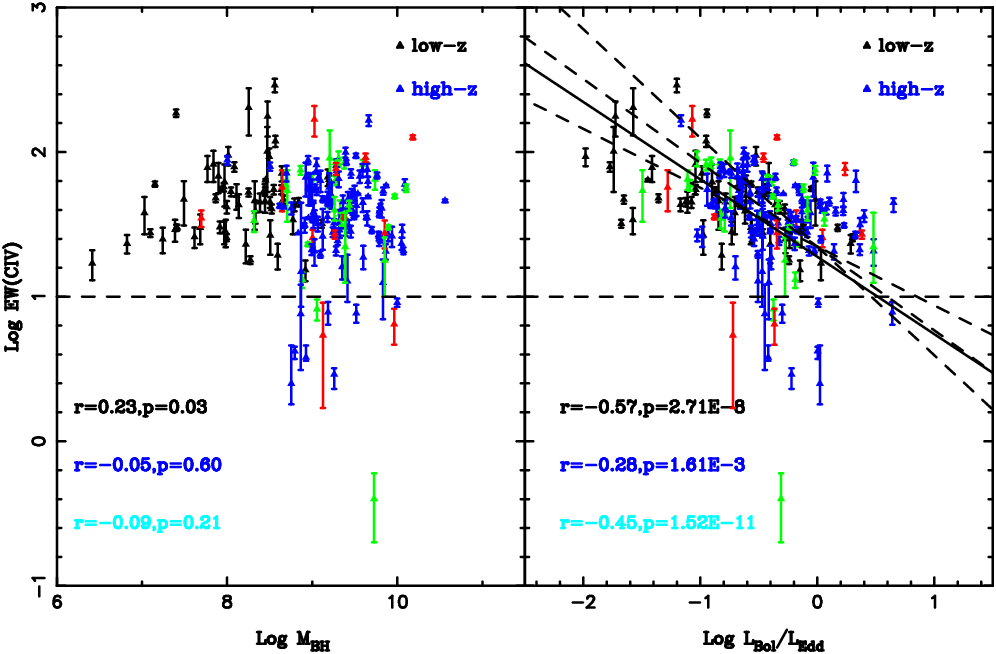}
\caption{Left: the relationship between \civ EW and \mbh. Right: the relationship between \civ EW and \leddR. The marks are the same as Fig. 10. The solid line is the best linear fitting of BCES bisector for the low-$z$ sample. The dotted lines are the best fitting of BCES bisector for the entire sample with $2\sigma$. BAL QSOs, RL QSOs and weak line QSOs are excluded from the entire sample during the fitting. }
\label{fig11}
\end{center}
\end{figure}

For low-$z$ PG QSOs, \cite{Baskin2004} found that the \civ Baldwin effect is weak. Fig. ~\ref{fig10} shows the \civ Baldwin effect for our low-$z$ and high-$z$ QSOs sample. For the 81 low-$z$ PG QSOs subsample, the \civ Baldwin effect is weak, with $r=-0.15$. It is consistent with the result by \cite{Baskin2004}. Considering the speciality of RL QSOs and BAL QSOs, we exclude 16 RL QSOs and 5 BAL QSOs, and find that the \civ Baldwin effect becomes stronger with $r=-0.25$. For high-$z$ QSOs subsample, the \civ Baldwin effect is stronger than low-$z$ PG QSOs subsample with $r=-0.34$. Excluding 12 RL QSOs, 7 BAL QSOs and 11 weak line QSOs in high-$z$ subsample, the \civ Baldwin effect becomes stronger with $r=-0.45$. For the total sample, we find there exists a \civ Baldwin effect with $r=-0.3$, which is consistent with the result by \cite{Bian2012} from 35019 QSOs in SDSS DR7. We notice that, for the total sample, excluding RL QSOs, BAL QSOs, and weak line QSOs, the correlation coefficient in \civ Baldwin effect doesn't increase.

For the high-$z$, we calculate the SMBH \mbh and \leddR using the same \hb emission line as that for low-$z$ PG QSOs. Fig. ~\ref{fig11} shows \civ EW versus \mbh (left-hand panel) and \leddR (right-hand panel). For low-$z$ PG QSOs subsample, there is a weak correlation between \civ EW and \mbh with $r=-0.23$ and $P_{\rm null}=0.03$. However, for high-$z$ QSOs subsample or the total sample, the correlations become weaker with $r=-0.05$, $P_{\rm null}=0.60$ and $r=-0.09$, $P_{\rm null}=0.21$, respectively. For low-$z$ PG QSOs subsample, \civ EW has a strong correlation with \leddR with $r=-0.57$, which is consistent with the result by \citep{Baskin2004}. For the high-$z$ subsample, the relation becomes weaker with r=-0.28. For the total sample, \civ EW has a strong correlation with $r=-0.45$, $P_{\rm null}=1.52\times 10^{-11}$. The correlation still exists when excluding RL QSOs, BAL QSOs and weak line QSOs. For the relation between \civ EW and \lbol (i.e., \lv), we find r=-0.15, -0.23, -0.27 for low-$z$, high-$z$, total sample, respectively. Considering that the correlation between \civ EW and \leddR is stronger than the relation between \civ EW and \mbh, and the \civ Baldwin effect, the Eddington ratio seems to be a better underlying physical parameter than the central SMBH \mbh in \civ Baldwin effect. Considering intrinsic scatter of 0.3 dex in \leddR, we  use the bivariate correlated errors and scatter method \citep[BCES;][]{Akritas1996} to perform the linear regression (Table.~\ref{table3}). The BCES Bisector best-fitting relation for low-$z$ PG QSOs subsample is,
\begin{eqnarray}
\rm log~EW(C~IV) = (-0.53\pm 0.10) log (\lv) \,\nonumber\\+ (1.28\pm 0.08)
\label{eq4}
\end{eqnarray}
is plotted as the solid line in Fig. ~\ref{fig11}. The high-$z$ QSOs follow the solid line found in low-$z$ PG QSOs subsample. Some of weak line QSOs deviated from the above fitting line. Considering the contamination in \civ EW measurement for BAL QSOs, and the jet effect in mass calculation for RL QSOs, we exclude RL QSOs, BAL QSOs and weak line QSOs, a new BCES Bisector best-fitting relation for the total QSOs sample is,
\begin{eqnarray}
\rm log~EW(C~IV) = (-0.58\pm 0.09) log (\lv) \,\nonumber\\+ (1.35\pm 0.05)
\label{eq5}
\end{eqnarray}
is plotted as the dash lines with $2 \sigma$ in the slope in Fig. ~\ref{fig11}. It is consistent with the result by \cite{Shemmer2015}. The slope of -0.56 is steeper than the slope of -0.268 found by \cite{Bian2012} with the \mgii-based \leddR. For RL QSOs, BAL QSOs and weak line QSOs, the \civ Baldwin effect, as well as the \leddR origin, need to be investigated in larger sample with reliable \civ EW, the UV continuum luminosity, \mbh and \leddR.

\section{Conclusions}
We compile a sample of 87 low-$z$ and 126 high-$z$ QSOs across the widest possible ranges of redshift ($0<z<5$) with available \hb and \civ observations. The \hb-based single-epoch SMBH \mbh and \leddR  are consistently calculated for QSOs from low-$z$ to high-$z$, which is used to investigate the underlying driver for the \civ Baldwin effect. The main conclusions can be summarized as follows.

(1) An empirical formula is presented to estimate the host correction in the continuum luminosity at 5100\AA. For low-$z$ PG QSOs, the estimated host fraction is less than 50\%. For 87 low-$z$ and 126 high-$z$ QSOs, the \hb-based single-epoch SMBH \mbh and \leddR are consistently calculated.

(2) Considering PC1/PC2 from optical PCA, it is found that PC1 has strong correlations with \mbh and \leddR. PC2 has a strong correlation with \mbh and a weak correlation with \leddR. It suggests that \leddR is the main driver of PC1, which is consistent with \cite{Boroson2002}. For low-$z$ PG QSOs, \civ EW has a relatively weak correlation with the continuum luminosity at 1350 \AA. The \civ EW has a strong correlation with the optical \feii strength or PC1.

(3) For low-$z$ PG QSOs subsample, excluding the RL QSOs and BAL QSOs, the \civ Baldwin effect becomes stronger. For low-$z$ and high-$z$ sample, there exists a \civ Baldwin effect with $r=-0.3$, which is consistent with the result by \cite{Bian2012} from 35019 QSOs in SDSS DR7. For the total sample, the correlation between \civ EW and \hb-based \mbh is very weak, and \civ EW has a strong correlation with \leddR. The Eddington ratio seems to be a better underlying physical parameter than the SMBH \mbh in \civ Baldwin effect. For RL QSOs, BAL QSOs and weak line QSOs, the \civ Baldwin effect, as well as the \leddR origin, need to be investigated in larger sample.

\section{ACKNOWLEDGEMENTS}
We are very grateful to the anonymous referee for her/his instructive comments which improved the content of the paper. This work has been supported by the National Science Foundations of China (Nos. 11373024, 11173016 and 11233003).

\begin{table*}
\label{table1}
\centering
\caption{The properties of 87 PG QSOs. Col.(1): Sequence number;
Col.(2): Name of PG QSOs, and the superscript of $^a$ in the Col. (2) indicates the reverberation mapping objects;
Col.(3): Redshift from \citep[]{Schmidt1983};
Col.(4): The continuum luminosity at 1350\AA\ in units of erg/s (in logarithm scale);
Col.(5): The equivalent widths of \civ in units of \AA;
Col.(6): The ratio of the equivalent widths of \feii (4434-4684\AA) and \hb;
Col.(7-8): The continuum luminosity at 5100 \AA\ \citep[]{Vestergaard2006,Neugebauer1987,Schmidt1983} and the host corrected luminosity at 5100\AA, in units of erg/s (in logarithm scale);
Col.(8): The host corrected continuum luminosity at 5100 \AA, in units of erg/s (in logarithm scale);
Col.(9): The \hb FWHM in units of \kms \citep[]{Vestergaard2006};
Col.(10, 11, 12): The host-corrected \mbh in this work (all in logarithm scale), the \mbh from other works \citep[]{Vestergaard2006,Baskin2004};
Col.(13): The host-corrected Eddington ratio (in logarithm scale);
Col.(14): The radio loudness (in logarithm scale);
Col.(15, 16): Eigenvector 1 and Eigenvector 2 of optical PCA.
}
\begin{lrbox}{\tablebox}
\begin{tabular}{l|l|l|l|l|l|l|l|l|l|l|l|r|r|r|r|}
\hline
N & Name  &  $z$ & $L_{1350}$  & EW(C IV) & $R_{\rm Fe II}$ & $L_{5100}$ &$L_{5100,cor}$&FWHM$_{\rm H\beta}$ & $M_{\rm BH,cor}$ & $M_{\rm BH,VP}$ & $M_{\rm BH, BL}$ & $L_{\rm Bol}/L_{\rm Edd}$  &  $R $ &$pc1$ & $pc2$  \\
  &       &      & (\ergs)       & (\AA) &                     & (\ergs)   & (\ergs)           & (\kms)       & ($M_{\odot})$  & ($M_{\odot})$  & ($M_{\odot}$)  & & & &  \\
(1) & (2)& (3)& (4)& (5)& (6)& (7)& (8)& (9)& (10)& (11)& (12)& (13) & (14) &(15)&(16)\\
\hline
1  &  0003+158     &0.450  & 46.050    & $   63.5\pm{  4.6}$ &0.00  &     46.018& 46.018&4750.7        &   9.273     & 9.273     &    9.055       &  -0.388          &   2.24    &  3.409     &  -0.796 \\
2  &$0003+199^{a}$ &0.025  & 44.232    & $   60.1\pm{  2.6}$ &0.62  &     44.160& 43.720&1585.0        &   7.152     & 7.152     &    7.220       &  -0.566          &  -0.57    & -1.896     &   1.860 \\
3  &  0007+106     &0.089  & 44.943    & $   59.0\pm{    5}$ &0.35  &     44.816& 44.734&5084.6        &   8.689     & 8.731     &    8.561       &  -1.089          &   2.29    &  2.302     &   0.188 \\
4  &$0026+129^{a}$ &0.142  & 45.240    & $   19.3\pm{  3.9}$ &0.51  &     45.100& 44.910&1821.0        &   8.594     & 8.594     &    7.833       &  -0.818          &   0.03    &  0.574     &  -0.248 \\
5  &  0043+039     &0.384  & 45.658    & $    5.4\pm{  3.7}$ &1.18  &     45.537& 45.537&5290.8        &   9.126     & 9.126     &    8.952       &  -0.722          &  -0.92    & -2.112     &  -2.484 \\
6  &  0049+171     &0.064  & 44.064    & $  203.0\pm{   73}$ &0.00  &     44.004& 43.813&5234.3        &   8.254     & 8.350     &    8.146       &  -1.575          &  -0.49    &  2.492     &   2.700 \\
7  &  0050+124     &0.061  & 44.755    & $   29.9\pm{  1.5}$ &1.47  &     44.794& 44.709&1171.4        &   7.402     & 7.444     &    7.238       &   0.173          &  -0.48    & -2.163     &  -0.627 \\
8  &$0052+251^{a}$ &0.155  & 45.277    & $  119.0\pm{ 10.5}$ &0.23  &     45.030& 44.750&5187.0        &   8.567     & 8.567     &    8.745       &  -0.951          &  -0.62    &  1.649     &   0.816 \\
9  &  0157+001     &0.164  & 45.099    & $   43.0\pm{    8}$ &0.71  &     44.975& 44.911&2431.9        &   8.137     & 8.169     &    8.006       &  -0.360          &   0.33    &  0.718     &   0.594 \\
10 &$0804+761^{a}$ &0.100  & 45.406    & $   45.0\pm{    3}$ &0.67  &     45.060& 44.850&3045.0        &   8.841     & 8.841     &    8.352       &  -1.125          &  -0.22    & -0.593     &  -0.651 \\
11 &  0838+770     &0.131  & 44.852    & $   50.0\pm{   10}$ &0.89  &     44.727& 44.634&2763.8        &   8.110     & 8.157     &    7.992       &  -0.610          &  -0.96    & -1.628     &  -0.060 \\
12 &$0844+349^{a}$ &0.064  & 44.634    & $   28.0\pm{    5}$ &0.89  &     44.490& 44.367&2386.0        &   7.966     & 7.966     &    7.759       &  -0.733          &  -1.52    & -2.101     &   0.629 \\
13 &$0921+525^{a}$ &0.035  & 43.758    & $  186.0\pm{   11}$ &0.14  &     43.630& 43.590&2079.0        &   7.400     & 7.400     &    7.206       &  -0.944          &   0.17    &  0.539     &   2.775 \\
14 &  0923+129     &0.190  & 43.923    & $   93.0\pm{   13}$ &0.53  &     43.860& 43.647&7598.4        &   8.495     & 8.601     &    7.233       &  -1.982          &  -0.85    & -0.756     &   1.260 \\
15 &  0923+201     &0.029  & 45.314    & $   28.0\pm{    6}$ &0.72  &     45.038& 44.981&1956.7        &   7.984     & 8.012     &    9.094       &  -0.136          &   0.32    & -0.129     &  -0.003 \\
16 &  0934+013     &0.050  & --        & $   -            $ &0.48  &     43.875& 43.664&1254.3        &   6.939     & 7.044     &    --          &  -0.409          &  -0.42    & -0.838     &   3.188 \\
17 &  0947+396     &0.206  & 44.975    & $   55.0\pm{    4}$ &0.23  &     44.808& 44.725&4816.7        &   8.638     & 8.679     &    8.530       &  -1.047          &  -0.60    &  0.981     &   1.244 \\
18 &$0953+414^{a}$ &0.239  & 45.646    & $   54.9\pm{    5}$ &0.25  &     45.400& 45.130&3111.0        &   8.441     & 8.441     &    8.488       &  -0.445          &  -0.36    &  0.800     &  -0.734 \\
19 &  1001+054     &0.161  & 44.979    & $   34.9\pm{  4.6}$ &0.82  &     44.741& 44.650&1699.8        &   7.696     & 7.741     &    7.645       &  -0.180          &  -0.30    & -2.452     &  -1.449 \\
20 &  1004+130     &0.240  & --        & $   -            $ &0.23  &     45.536& 45.536&6290.4        &   9.275     & 9.275     &    --          &  -0.873          &   2.36    &  0.818     &  -3.485 \\
21 &  1011-040     &0.058  & 44.398    & $   25.0\pm{    5}$ &0.73  &     44.259& 44.105&1381.0        &   7.243     & 7.320     &    7.190       &  -0.272          &  -1.00    & -2.225     &   0.682 \\
22 &  1012+008     &0.185  & 45.104    & $   23.0\pm{    6}$ &0.66  &     45.011& 44.951&2614.7        &   8.220     & 8.250     &    8.069       &  -0.403          &  -0.30    & -0.145     &  -0.549 \\
23 &  1022+519     &0.045  & 43.694    & $   38.0\pm{   11}$ &1.08  &     43.696& 43.456&1566.4        &   7.028     & 7.148     &    6.940       &  -0.706          &  -0.64    & -3.002     &   2.245 \\
24 &  1048-090     &0.344  & 45.697    & $   91.0\pm{   50}$ &0.09  &     45.596& 45.596&5610.8        &   9.206     & 9.206     &    9.022       &  -0.744          &  -1.00    &  2.683     &  -1.268 \\
25 &  1048+342     &0.167  & 44.908    & $   46.0\pm{   17}$ &0.32  &     44.708& 44.613&3580.9        &   8.324     & 8.372     &    8.241       &  -0.845          &   2.58    &  0.559     &  -0.192 \\
26 &  1049-005     &0.357  & 45.713    & $   67.0\pm{  8.8}$ &0.56  &     45.633& 45.633&5350.6        &   9.183     & 9.183     &    8.989       &  -0.684          &  -0.60    &  2.174     &  -0.323 \\
27 &  1100+772     &0.313  & 45.717    & $   84.0\pm{  4.9}$ &0.21  &     45.575& 45.575&6151.2        &   9.275     & 9.275     &    9.112       &  -0.834          &   2.51    &  3.756     &  -0.652 \\
28 &  1103-006     &0.425  & 45.748    & $   37.2\pm{    9}$ &0.60  &     45.667& 45.667&6182.6        &   9.326     & 9.326     &    9.132       &  -0.793          &   2.43    &  1.681     &  -1.915 \\
29 &  1114+445     &0.144  & 44.842    & $   55.0\pm{  4.1}$ &0.20  &     44.734& 44.642&4554.4        &   8.548     & 8.594     &    8.415       &  -1.040          &  -0.89    & -0.065     &  -0.203 \\
30 &  1115+407     &0.154  & 44.720    & $   25.9\pm{  4.2}$ &0.54  &     44.619& 44.513&1678.8        &   7.616     & 7.669     &    7.505       &  -0.237          &  -0.77    & -1.291     &  -0.116 \\
31 &  1116+215     &0.177  & 45.641    & $   40.5\pm{  2.9}$ &0.47  &     45.397& 45.378&2896.9        &   8.523     & 8.532     &    8.425       &  -0.279          &  -0.14    &  0.006     &  -1.223 \\
32 &  1119+120     &0.049  & 44.172    & $   29.0\pm{    5}$ &0.90  &     44.132& 43.960&1772.9        &   7.387     & 7.473     &    7.280       &  -0.561          &  -0.82    & -1.677     &   1.584 \\
33 &  1121+422     &0.234  & 44.979    & $   41.7\pm{  4.1}$ &0.37  &     44.883& 44.808&2192.3        &   7.996     & 8.033     &    7.856       &  -0.322          &  -1.00    & -0.921     &   0.710 \\
34 &  1126-041     &0.060  & 44.517    & $   30.0\pm{    7}$ &1.07  &     44.385& 44.248&2111.1        &   7.683     & 7.752     &    7.598       &  -0.569          &  -0.77    & -2.181     &  -0.279 \\
35 &  1149-110     &0.049  & 44.167    & $   82.0\pm{   20}$ &0.36  &     44.107& 43.931&3032.2        &   7.839     & 7.927     &    7.729       &  -1.042          &  -0.06    &  0.001     &   2.565 \\
36 &  1151+117     &0.176  & 44.988    & $   26.6\pm{  7.1}$ &0.24  &     44.756& 44.666&4284.3        &   8.507     & 8.552     &    8.435       &  -0.975          &  -1.15    & -0.355     &  -0.321 \\
37 &  1202+281     &0.165  & 44.763    & $  290.0\pm{ 31.3}$ &0.29  &     44.601& 44.492&5036.4        &   8.560     & 8.615     &    8.462       &  -1.202          &  -0.72    &  1.745     &   0.468 \\
38 &  1211+143     &0.085  & 45.236    & $   55.7\pm{  1.8}$ &0.52  &     45.071& 45.018&1816.9        &   7.938     & 7.964     &    7.831       &  -0.053          &  -0.87    & -0.302     &  -0.316 \\
39 &  1216+069     &0.334  & 45.700    & $   64.5\pm{  4.4}$ &0.20  &     45.721& 45.721&5179.9        &   9.199     & 9.199     &    8.954       &  -0.612          &   0.22    &  1.117     &  -1.401 \\
40 &$1226+023^{a}$ &0.158  & 46.218    & $   23.0\pm{  0.7}$ &0.57  &     46.020& 45.900&3500.0        &   8.947     & 8.947     &    8.876       &  -0.181          &   3.06    &  1.193     &  -2.548 \\
41 &$1229+204^{a}$ &0.064  & 44.554    & $   48.0\pm{    3}$ &0.59  &     44.390& 43.640&3335.0        &   7.865     & 7.865     &    8.004       &  -1.359          &  -0.96    & -0.861     &   0.804 \\
42 &  1244+026     &0.048  & 44.204    & $   17.0\pm{    4}$ &1.20  &     43.801& 43.578& 720.6        &   6.414     & 6.526     &    6.614       &   0.030          &  -0.28    & -3.500     &   2.584 \\
43 &  1259+593     &0.472  & 46.007    & $   15.3\pm{  2.5}$ &1.27  &     45.906& 45.906&3377.3        &   8.920     & 8.920     &    8.738       &  -0.148          &  -1.00    & -2.105     &  -2.117 \\
44 &  1302-102     &0.286  & 46.023    & $   13.1\pm{  1.6}$ &0.19  &     45.827& 45.827&3383.4        &   8.882     & 8.882     &    8.749       &  -0.189          &   2.27    &  1.885     &  -1.579 \\
45 &$1307+085^{a}$ &0.155  & 45.244    & $   71.2\pm{  8.5}$ &0.19  &     45.010& 44.790&5307.0        &   8.643     & 8.643     &    8.541       &  -0.987          &  -1.00    &  1.384     &   0.342 \\
46 &  1309+355     &0.184  & 45.088    & $   33.5\pm{  5.5}$ &0.28  &     45.014& 44.954&2917.3        &   8.317     & 8.347     &    8.155       &  -0.497          &   1.26    &  1.441     &  -0.906 \\
47 &  1310-108     &0.035  & 43.930    & $   78.0\pm{   16}$ &0.38  &     43.725& 43.490&3606.0        &   7.769     & 7.887     &    7.759       &  -1.413          &  -1.00    &  0.751     &   4.280 \\
48 &  1322+659     &0.168  & 45.020    & $   52.6\pm{  3.4}$ &0.59  &     44.980& 44.917&2765.4        &   8.252     & 8.284     &    8.076       &  -0.469          &  -0.92    & -0.992     &   0.850 \\
49 &  1341+258     &0.087  & 44.477    & $   62.0\pm{   20}$ &0.38  &     44.344& 44.202&3013.9        &   7.969     & 8.040     &    7.878       &  -0.901          &  -0.92    & -0.742     &   1.268 \\
50 &  1351+236     &0.055  & 43.822    & $  101.0\pm{   48}$ &1.18  &     44.048& 43.864&6527.2        &   8.471     & 8.563     &    8.216       &  -1.741          &  -0.59    & -0.702     &   0.547 \\
51 &  1351+640     &0.087  & 44.953    & $   43.3\pm{  4.4}$ &0.24  &     44.835& 44.755&5646.1        &   8.791     & 8.831     &    8.656       &  -1.170          &   0.64    &  1.443     &   0.897 \\
52 &  1352+183     &0.158  & 45.024    & $   45.1\pm{  6.5}$ &0.46  &     44.816& 44.734&3580.6        &   8.385     & 8.426     &    8.299       &  -0.785          &  -0.96    & -0.381     &   0.106 \\
53 &  1354+213     &0.300  & --        & $   -            $ &0.31  &     44.977& 44.913&4126.7        &   8.598     & 8.630     &    --          &  -0.818          &  -1.10    &  1.282     &   0.375 \\
54 &  1402+261     &0.164  & 45.217    & $   30.3\pm{  2.8}$ &1.23  &     44.983& 44.920&1873.7        &   7.915     & 7.947     &    7.845       &  -0.129          &  -0.64    & -2.466     &  -0.714 \\
55 &  1404+226     &0.098  & 44.299    & $   23.3\pm{  3.4}$ &1.01  &     44.379& 44.242& 787.3        &   6.823     & 6.892     &    6.713       &   0.285          &  -0.33    & -3.123     &  -0.929 \\
56 &$1411+442^{a}$ &0.089  & 44.693    & $   56.9\pm{   18}$ &0.49  &     44.620& 44.500&2640.0        &   8.646     & 8.646     &    7.874       &  -1.280          &  -0.89    & -1.150     &  -0.365 \\
57 &  1415+451     &0.114  & 44.572    & $   57.3\pm{  3.9}$ &1.25  &     44.561& 44.447&2591.2        &   7.961     & 8.018     &    7.797       &  -0.647          &  -0.77    & -3.045     &  -0.211 \\
58 &  1416-129     &0.129  & 45.510    & $  168.1\pm{ 40.2}$ &0.18  &     45.135& 45.089&6098.0        &   9.025     & 9.048     &    9.002       &  -1.070          &   0.06    &  2.186     &  -0.667 \\
59 &  1425+267     &0.366  & 45.392    & $   64.8\pm{   10}$ &0.11  &     45.761& 45.761&9404.7        &   9.737     & 9.737     &    9.317       &  -1.110          &   1.73    &  3.763     &  -1.448 \\
60 &$1426+015^{a}$ &0.086  & 45.158    & $   32.0\pm{    2}$ &0.39  &     44.880& 44.570&6808.0        &   9.113     & 9.113     &    8.921       &  -1.677          &  -0.55    &  0.289     &   0.434 \\

\hline
\end{tabular}
\end{lrbox}
\scalebox{0.8}{\usebox{\tablebox}}
\end{table*}

\setcounter{table}{0}
\begin{table*}
\label{table2}
\centering
\centering \caption{-- continue}
\begin{lrbox}{\tablebox}
\begin{tabular}{l|l|l|l|l|l|l|l|l|l|l|l|l|l|l|l|}
\hline
N & Name  &  $z$ & $L_{1350}$  & EW(C IV) & $R_{\rm Fe II}$ & $L_{5100}$ &$L_{5100,cor}$&FWHM$_{\rm H\beta}$ & $M_{\rm BH,cor}$ & $M_{\rm BH,VP}$ & $M_{\rm BH, BL}$ & $L_{\rm Bol}/L_{\rm Edd}$  &  $R $ &$pc1$ & $pc2$  \\
(1) & (2)& (3)& (4)& (5)& (6)& (7)& (8)& (9)& (10)& (11)& (12)& (13) & (14) &(15)&(16)\\
\hline
61 &  1427+480     &0.221  & 44.988    & $   53.2\pm{  3.7}$ &0.36  &     44.759& 44.670&2515.3        &   8.046     & 8.091     &    7.978       &  -0.510          &  -0.80    &  0.770     &   0.892 \\
62 &  1435-067     &0.129  & 45.242    & $   39.0\pm{    7}$ &0.45  &     44.918& 44.847&3156.9        &   8.332     & 8.368     &    8.300       &  -0.619          &  -1.15    & -0.732     &  -0.160 \\
63 &  1440+356     &0.077  & 44.676    & $   30.1\pm{  1.4}$ &1.19  &     44.546& 44.430&1393.5        &   7.413     & 7.471     &    7.335       &  -0.117          &  -0.43    & -1.874     &   0.401 \\
64 &  1444+407     &0.267  & 45.390    & $   17.9\pm{  1.1}$ &1.45  &     45.203& 45.164&2456.5        &   8.273     & 8.292     &    8.158       &  -0.242          &  -1.10    & -2.562     &  -1.225 \\
65 &  1448+273     &0.065  & --        & $   -            $ &0.90  &     44.482& 44.358& 814.7        &   6.911     & 6.973     &    --          &   0.313          &  -0.60    & -1.876     &   0.758 \\
66 &  1501+106     &0.036  & 44.664    & $   64.0\pm{    1}$ &0.35  &     44.285& 44.135&5454.1        &   8.451     & 8.526     &    8.482       &  -1.450          &  -0.44    &  1.255     &   1.827 \\
67 &  1512+370     &0.371  & 45.656    & $   84.3\pm{  7.2}$ &0.00  &     45.602& 45.602&6802.7        &   9.376     & 9.376     &    9.168       &  -0.908          &   2.28    &  3.934     &  -0.923 \\
68 &  1519+226     &0.137  & 44.820    & $   68.0\pm{   16}$ &1.01  &     44.710& 44.615&2187.3        &   7.897     & 7.945     &    7.777       &  -0.416          &  -0.05    & -2.021     &  -0.221 \\
69 &  1534+580     &0.030  & 43.836    & $   79.0\pm{    6}$ &0.27  &     43.687& 43.445&5323.5        &   8.085     & 8.206     &    8.047       &  -1.774          &  -0.15    &  1.690     &   4.678 \\
70 &  1535+547     &0.038  & 43.991    & $   27.6\pm{  1.7}$ &0.47  &     43.961& 43.764&1420.4        &   7.097     & 7.195     &    7.010       &  -0.467          &  -0.85    & -2.801     &   0.279 \\
71 &  1543+489     &0.400  & 45.567    & $   25.6\pm{  1.4}$ &0.86  &     45.445& 45.431&1529.2        &   7.995     & 8.001     &    7.844       &   0.303          &  -0.82    & -2.791     &  -1.797 \\
72 &  1545+210     &0.266  & 45.594    & $   90.5\pm{ 10.5}$ &0.00  &     45.428& 45.413&7021.7        &   9.309     & 9.317     &    9.165       &  -1.030          &   2.62    &  3.654     &  -1.141 \\
73 &  1552+085     &0.119  & 44.758    & $   47.0\pm{   16}$ &1.02  &     44.704& 44.608&1377.0        &   7.492     & 7.540     &    7.364       &  -0.018          &  -0.35    & -2.872     &  -0.730 \\
74 &  1612+261     &0.131  & 44.872    & $   94.6\pm{ 13.9}$ &0.18  &     44.717& 44.623&2490.9        &   8.014     & 8.061     &    7.913       &  -0.525          &   0.45    &  2.143     &   1.693 \\
75 &$1613+658^{a}$ &0.129  & 44.849    & $   54.0\pm{    3}$ &0.38  &     44.840& 44.710&8441.0        &   8.446     & 8.446     &    8.953       &  -0.870          &   0.00    &  2.061     &   0.205 \\
76 &$1617+175^{a}$ &0.114  & 45.203    & $   34.0\pm{    7}$ &0.60  &     44.850& 44.330&5316.0        &   8.774     & 8.774     &    8.729       &  -1.578          &  -0.14    & -0.199     &  -0.758 \\
77 &  1626+554     &0.133  & 44.785    & $   45.6\pm{  7.6}$ &0.32  &     44.580& 44.469&4473.8        &   8.446     & 8.501     &    8.371       &  -1.111          &  -0.96    &  0.127     &   0.720 \\
78 &$1700+518^{a}$ &0.292  & --        & $   -            $ &1.42  &     45.680& 45.530&2185.0        &   8.893     & 8.893     &    --          &  -0.497          &   0.37    & -3.505     &  -3.873 \\
79 &  1704+608     &0.371  & 45.779    & $   34.8\pm{  5.2}$ &0.00  &     45.702& 45.702&6552.4        &   9.394     & 9.394     &    9.198       &  -0.826          &   2.81    &  5.905     &  -0.594 \\
80 &  2112+059     &0.466  & 46.304    & $   25.5\pm{  3.5}$ &0.63  &     46.181& 46.181&3176.4        &   9.004     & 9.004     &    8.834       &   0.043          &  -0.49    & -0.988     &  -2.688 \\
81 &$2130+099^{a}$ &0.061  & 44.792    & $   47.0\pm{    3}$ &0.64  &     44.540& 44.140&2294.0        &   8.660     & 8.660     &    7.805       &  -1.654          &  -0.49    & -0.487     &   0.919 \\
82 &  2209+184     &0.070  & 44.601    & $   54.0\pm{   21}$ &0.44  &     44.469& 44.344&6487.5        &   8.706     & 8.769     &    8.601       &  -1.496          &   2.15    &  0.570     &  -0.599 \\
83 &  2214+139     &0.067  & 44.635    & $   45.0\pm{    4}$ &0.32  &     44.662& 44.561&4532.0        &   8.503     & 8.554     &    8.308       &  -1.076          &  -1.30    & -1.662     &  -1.135 \\
84 &  2233+134     &0.325  & --        & $   -            $ &0.89  &     45.327& 45.301&1709.2        &   8.026     & 8.039     &    --          &   0.141          &  -0.55    & -1.320     &  -1.086 \\
85 &  2251+113     &0.323  & 45.807    & $   66.0\pm{  3.5}$ &0.32  &     45.692& 45.692&4147.2        &   8.992     & 8.992     &    8.816       &  -0.433          &   2.56    &  1.857     &  -2.137 \\
86 &  2304+042     &0.042  & 44.040    & $  176.0\pm{   48}$ &0.09  &     44.066& 43.884&6486.8        &   8.476     & 8.567     &    8.320       &  -1.726          &  -0.60    &  0.467     &   2.736 \\
87 &  2308+098     &0.432  & 45.789    & $   81.5\pm{  6.8}$ &0.00  &     45.777& 45.777&7914.3        &   9.595     & 9.595     &    9.372       &  -0.952          &   2.27    &  3.539     &  -1.259 \\

\hline
\end{tabular}
\end{lrbox}
\scalebox{0.8}{\usebox{\tablebox}}
\end{table*}

\begin{table*}
\centering
\label{table2}
\caption{The properties of high-z QSOs. Col.(1): Sequence number;
Col.(2): Name of high-$z$ QSOs;
Col.(3): Redshift;
Col.(4): The continuum luminosity at 1350 \AA in units of \ergs (in logarithm scale);
Col.(5): The equivalent widths of \civ in units of \AA;
Col.(6): The continuum luminosity at 5100\AA in units of \ergs (in logarithm scale);
Col.(7): The \ha and \hb FWHM in units of \kms;
Col.(8): Reference. A11: \citep[]{Assef2011}, H12: \citep[]{Ho2012}, J15: \citep[]{Jun2015}, N07: \citep[]{Netzer2007}, S04: \citep[]{Shemmer2004}, S15: \citep[]{Shemmer2015}, Sy12: \citep[]{Shen2012}. The number of objects in each literature is 7, 1, 26,15, 10, 7, 60 respectively.
Col.(9): The single-epoch \hb-based SMBH mass;
Col.(10): The Eddington ratio;
Col.(11): The radio loudness in logarithm scale \citep[]{Shen2011}.
}
\begin{lrbox}{\tablebox}
\begin{tabular}{l|l|l|l|l|l|l|l|l|r|r|l}
\hline
N & Name  &  $z$   & $L_{1350}$  &EW(C IV) &$L_{5100}$& FWHM($\ha,\hb$) & ref. &$\mbh$ & $L_{Bol}/L_{Edd}$  & $R$ \\
&         &        &  (\ergs)    & (\AA)   & (\ergs)  & (\kms) &     & ($M_{\odot}$) & & \\
(1) & (2)& (3)& (4)& (5)& (6)& (7)& (8)& (9)& (10) &(11) \\
\hline
1    & SDSS225800.02-084143    &   1.494    &         46.585  &  $  27.479 $          &    45.84  &    (  --, 3188  )   &    sy12 &  8.835     &     -0.133  &  --           \\
2    & SDSS035856.73-054023    &   1.506    &         46.273  &  $  31.912 \pm{3.737 } $          &    45.80    &    (  --, 4240  )   &    sy12 &  9.065     &     -0.398  &  --           \\
3    & SDSS081331.28+254503    &   1.51     &         47.169  &  $  42.45  \pm{1.396 } $          &    46.96  &    (  --, 5091  )   &    sy12 &  9.802     &     0.021   &  --           \\
4    & HS0810+2554             &   1.51     &         47.169  &  $  42.45  \pm{1.396 } $          &    44.84   &    (  --, 4400  )   &    A11  &  8.617     &     -0.911  &  --           \\
5    & SDSS133321.90+005824    &   1.514    &         46.455  &  $  99.303 \pm{7.916 } $          &    45.90  &    (  --, 5841  )   &    sy12 &  9.391     &     -0.628  &  --           \\
6    & SDSS152111.86+470539    &   1.516    &         46.475  &  $  65.392 \pm{8.13  } $          &    45.97  &    (  --, 5100  )   &    sy12 &  9.312     &     -0.472  &  --           \\
7    & FBQ1633+3134            &   1.518    &         46.677  &  $  19.631 \pm{1.034 } $          &    45.72   &    (  --, 4600  )   &    A11  &  9.096     &     -0.509  &  --           \\
8    & SDSS085543.26+002908    &   1.523    &         46.109  &  $  80.674 \pm{8.211 } $          &    45.78  &    (  --, 5518  )   &    sy12 &  9.284     &     -0.637  &  --           \\
9    & SDSS123355.21+031327    &   1.526    &         46.329  &  $  74.82  \pm{7.72  } $          &    45.93  &    (  --, 7227  )   &    sy12 &  9.591     &     -0.798  &  --           \\
10   & SDSS104910.31+143227    &   1.536    &         46.299  &  $  56.828 \pm{2.845 } $          &    46.01  &    (  --, 3629  )   &    sy12 &  9.036     &     -0.157  &  --           \\
11   & SDSS143230.57+012435    &   1.538    &         46.513  &  $  40.896 \pm{3.845 } $          &    45.97  &    (  --, 2699  )   &    sy12 &  8.755     &     0.077   &  --           \\
12   & SDSS154212.90+111226    &   1.538    &         46.635  &  $  52.345 \pm{6.235 } $          &    46.06  &    (  --, 6028  )   &    sy12 &  9.498     &     -0.577  &  --           \\
13   & SDSSJ105023.68-01055    &   1.539    &         45.69   &  $  7.607  \pm{4.494 } $          &    45.55   &    (  3891, 4798)   &    H12  &  8.865     &     -0.449  &  --           \\
14   & SDSS081344.15+152221    &   1.541    &         46.537  &  $  30.334 \pm{3.931 } $          &    46.03  &    (  --, 5403  )   &    sy12 &  9.391     &     -0.494  &  --           \\
15   & SDSS082146.22+571226    &   1.542    &         46.75   &  $  30.836 \pm{3.051 } $          &    46.31  &    (  --, 4804  )   &    sy12 &  9.429     &     -0.251  &  --           \\
16   & SDSS074029.82+281458    &   1.543    &         46.495  &  $  94.255 \pm{3.388 } $          &    46.04  &    (  --, 6190  )   &    sy12 &  9.514     &     -0.607  &  --           \\
17   & SDSS015733.87-004824    &   1.545    &         46.358  &  $  16.088 \pm{2.94  } $          &    45.79  &    (  --, 5953  )   &    sy12 &  9.352     &     -0.7    &  --           \\
18   & SDSS123442.16+052126    &   1.549    &         46.565  &  $  40.881 \pm{5.644 } $          &    46.16  &    (  --, 8173  )   &    sy12 &  9.816     &     -0.787  &  --           \\
19   & SDSS101447.54+521320    &   1.55     &         46.55   &  $  38.6   \pm{5.77  } $          &    46.02  &    (  --, 3534  )   &    sy12 &  9.015     &     -0.132  &  --           \\
20   & SDSS135439.70+301649    &   1.55     &         46.31   &  $  49.973 \pm{5.56  } $          &    46.10  &    (  --, 5666  )   &    sy12 &  9.465     &     -0.502  &  --           \\
21   & SDSS171030.20+602347    &   1.552    &         46.453  &  $  62.286 \pm{5.034 } $          &    46.13  &    (  --, 6924  )   &    sy12 &  9.655     &     -0.66   &  --           \\
22   & SDSS223246.80+134702    &   1.555    &         46.423  &  $  68.885 \pm{3.261 } $          &    46.06   &    (  --, 8146  )   &    sy12 &  9.762     &     -0.836  &  --           \\
23   & SDSS100930.51+023052    &   1.556    &         45.985  &  $  81.689 \pm{8.522 } $          &    45.59  &    (  --, 4721  )   &    sy12 &  9.051     &     -0.599  &  --           \\
24   & SDSS124006.70+474003    &   1.559    &         46.279  &  $  74.844 \pm{5.32  } $          &    45.98  &    (  --, 3038  )   &    sy12 &  8.864     &     -0.020   &  41.14        \\
25   & SDSS093318.49+141340    &   1.562    &         46.388  &  $  75.518 \pm{7.676 } $          &    46.10  &    (  --, 6992  )   &    sy12 &  9.649     &     -0.683  &  --           \\
26   & SDSS113829.33+040101    &   1.564    &         46.738  &  $  26.227 \pm{4.187 } $          &    46.09  &    (  --, 9814  )   &    sy12 &  9.937     &     -0.984  &  --           \\
27   & SDSS084451.91+282607    &   1.57     &         46.314  &  $  71.545 \pm{9.017 } $          &    45.92  &    (  --, 2599  )   &    sy12 &  8.698     &     0.085   &  --           \\
28   & SDSS094126.49+044328    &   1.571    &         46.291  &  $  71.17  \pm{4.79  } $          &    45.85  &    (  --, 6441  )   &    sy12 &  9.455     &     -0.735  &  --           \\
29   & SDSS081227.19+075732    &   1.575    &         46.539  &  $  26.639 \pm{2.516 } $          &    46.00  &    (  --, 9813  )   &    sy12 &  9.895     &     -1.026  &  --           \\
30   & SDSS204538.96-005115    &   1.589    &         46.324  &  $  48.858 \pm{4.253 } $          &    45.84  &    (  --, 4721  )   &    sy12 &  9.18      &     -0.470   &  --           \\
31   & SDSS014705.42+133210    &   1.59     &         46.718  &  $  32.601 \pm{7.143 } $          &    46.21  &    (  --, 4506  )   &    sy12 &  9.321     &     -0.248  &  --           \\
32   & SDSS091754.44+043652    &   1.59     &         46.132  &  $  165.696\pm{14.534} $          &    45.63  &    (  --, 9310  )   &    sy12 &  9.664     &     -1.166  &  --           \\
33   & SDSS114023.40+301651    &   1.592    &         46.891  &  $  47.193 \pm{5.803 } $          &    46.42  &    (  --, 4463  )   &    sy12 &  9.421     &     -0.132  &  --           \\
34   & SDSS125140.82+080718    &   1.596    &         46.747  &  $  37.971 \pm{8.129 } $          &    46.09  &    (  --, 3389  )   &    sy12 &  9.016     &     -0.058  &  --           \\
35   & SDSS104603.22+112828    &   1.602    &         46.234  &  $  56.916 \pm{14.337} $          &    45.87  &    (  --, 4768  )   &    sy12 &  9.2       &     -0.467  &  --           \\
36   & SDSS204009.62-065402    &   1.61     &         45.957  &  $  54.143 \pm{10.508} $          &    45.46   &    (  --, 4770  )   &    sy12 &  8.997     &     -0.671  &  --           \\
37   & SDSS155240.40+194816    &   1.611    &         46.537  &  $  35.068 \pm{4.003 } $          &    46.04  &    (  --, 7202  )   &    sy12 &  9.647     &     -0.737  &  --           \\
38   & SDSS083850.15+261105    &   1.612    &         47.136  &  $  39.073 \pm{0.988 } $          &    46.60  &    (  --, 4038  )   &    sy12 &  9.42      &     0.041   &  --           \\
39   & SDSS002948.04-095639    &   1.616    &         46.324  &  $  66.829 \pm{17.563} $          &    46.02  &    (  --, 3171  )   &    sy12 &  8.923     &     -0.035  &  --           \\
40   & SDSS135023.68+265243    &   1.617    &         46.677  &  $  44.182 \pm{2.062 } $          &    46.22  &    (  --, 3813  )   &    sy12 &  9.182     &     -0.097  &  --           \\
41   & SDSS205554.08+004311    &   1.618    &         46.143  &  $  57.207 \pm{14.318} $          &    45.45  &    (  --, 4589  )   &    sy12 &  8.957     &     -0.643  &  --           \\
42   & SDSS111949.30+233249    &   1.62     &         46.21   &  $  67.852 \pm{6.652 } $          &    46.07  &    (  --, 6416  )   &    sy12 &  9.558     &     -0.625  &  --           \\
43   & SDSS004149.64-094705    &   1.622    &         46.843  &  $  35.272 \pm{1.085 } $          &    46.19  &    (  --, 6552  )   &    sy12 &  9.636     &     -0.583  &  --           \\
44   & SDSS160456.14-001907    &   1.629    &         46.715  &  $  42.693 \pm{3.359 } $          &    46.23  &    (  --, 5064  )   &    sy12 &  9.435     &     -0.337  &  1843.61      \\
45   & SDSS141949.39+060654    &   1.638    &         46.666  &  $  28.927 \pm{2.503 } $          &    45.94  &    (  --, 5252  )   &    sy12 &  9.319     &     -0.516  &  --           \\
46   & SDSS100401.27+423123    &   1.653    &         47.003  &  $  20.783 \pm{1.471 } $          &    46.34  &    (  --, 3977  )   &    sy12 &  9.28      &     -0.072  &  --           \\
47   & SDSS142841.97+592552    &   1.653    &         46.482  &  $  35.699 \pm{2.802 } $          &    46.09  &    (  --, 4095  )   &    sy12 &  9.179     &     -0.224  &  --           \\
48   & SDSS020044.50+122319    &   1.656    &         46.607  &  $  28.453 \pm{1.061 } $          &    45.99  &    (  --, 4759  )   &    sy12 &  9.26      &     -0.404  &  --           \\
49   & SDSS204536.56-010147    &   1.658    &         47.112  &  $  37.154 \pm{1.789 } $          &    46.52  &    (  --, 6458  )   &    sy12 &  9.789     &     -0.405  &  --           \\
50   & SDSS110240.16+394730    &   1.659    &         46.397  &  $  34.553 \pm{4.662 } $          &    45.95  &    (  --, 5289  )   &    sy12 &  9.333     &     -0.514  &  --           \\
51   & SDSSJ094533.98+10095    &   1.662    &         46.46   &  $  2.9    ^{+0.3   }_{-0.6   } $ &    46.17   &    (  --, 4278  )   &    S15  &  9.257     &     -0.221  &  --           \\
52   & SDSS094913.05+175155    &   1.666    &         46.668  &  $  36.279 \pm{2.006 } $          &    46.18   &    (  --, 5581  )   &    sy12 &  9.493     &     -0.447  &  --           \\
53   & SDSS213748.44+001220    &   1.668    &         46.535  &  $  8.204  \pm{1.359 } $          &    45.82  &    (  --, 4163  )   &    sy12 &  9.056     &     -0.375  &  192.41       \\
54   & SDSS153859.45+053705    &   1.681    &         46.508  &  $  34.223 \pm{1.528 } $          &    46.06  &    (  --, 3414  )   &    sy12 &  9.004     &     -0.083  &  --           \\
55   & SDSS162103.98+002905    &   1.681    &         46.271  &  $  28.291 \pm{2.598 } $          &    46.02   &    (  --, 5835  )   &    sy12 &  9.452     &     -0.566  &  --           \\
56   & SDSS041255.16-061210    &   1.684    &         46.763  &  $  32.237 \pm{2.201 } $          &    46.05  &    (  --, 4566  )   &    sy12 &  9.256     &     -0.336  &  --           \\
57   & SDSS0246-0825           &   1.686    &         46.098  &  $  87.842 \pm{5.921 } $          &    44.59   &    (  --, 2500  )   &    A11  &  8.001     &     -0.545  &  --           \\
58   & SDSS105951.05+090905    &   1.688    &         46.797  &  $  84.549 \pm{3.086 } $          &    46.33  &    (  --, 4605  )   &    sy12 &  9.402     &     -0.205  &  --           \\
59   & SDSS112542.29+000101    &   1.689    &         46.495  &  $  84.899 \pm{2.669 } $          &    46.23  &    (  --, 4321  )   &    sy12 &  9.298     &     -0.198  &  79.93        \\
60   & SDSS101504.75+123022    &   1.69     &         46.632  &  $  24.331 \pm{1.592 } $          &    46.11  &    (  --, 4654  )   &    sy12 &  9.298     &     -0.327  &  --           \\
61   & SDSSJ141730.92+07332    &   1.704    &         46.304  &  $  2.5    ^{+2.1   }_{-0.7   } $ &    45.91   &    (  --, 2784  )   &    S15  &  8.754     &     0.022   &  --           \\

\hline
\end{tabular}
\end{lrbox}
\scalebox{0.9}{\usebox{\tablebox}}
\end{table*}

\setcounter{table}{1}
\begin{table*}
\centering \caption{-- continue}
\begin{lrbox}{\tablebox}
\begin{tabular}{l|l|l|l|l|l|l|l|l|r|r|l}
\hline
N & Name  &  $z$   & $L_{1350}$  &EW(C IV) &$L_{5100}$& FWHM($\ha,\hb$) & ref. &$\mbh$ & $L_{Bol}/L_{Edd}$  & $R$ \\
&         &        &  (\ergs)    & (\AA)   & (\ergs)  & (\kms) &     & ($M_{\odot}$) & & \\
(1) & (2)& (3)& (4)& (5)& (6)& (7)& (8)& (9)& (10) &(11) \\
\hline
62   & PG1115+080              &   1.735    &         47.303  &  $  41.557 \pm{0.743 } $          &    44.93   &    (  --, 4400  )   &    A11  &  8.662     &     -0.866  &  --           \\
63   & SDSSJ083650.86+14253    &   1.745    &         46.112  &  $  4.2    ^{+0.3   }_{-0.5   } $ &    45.93   &    (  --, 2880  )   &    S15  &  8.794     &     0.002   &  --           \\
64   & SDSSJ141141.96+14023    &   1.745    &         46.238  &  $  3.8    ^{+0.8   }_{-0.2   } $ &    45.64   &    (  --, 3966  )   &    S15  &  8.927     &     -0.420   &  --           \\
65   & SDSS122039.45+000427    &   2.047    &         46.838  &  $  45.22  \pm{1.656 } $          &    46.39  &    (  --, 3651  )   &    sy12 &  9.23      &     0.027   &  --           \\
66   & SDSS143645.80+633637    &   2.068    &         47.003  &  $  49.388 \pm{1.512 } $          &    46.72  &    (  --, 7056  )   &    sy12 &  9.969     &     -0.379  &  1670.64      \\
67   & SDSS014944.43+150106    &   2.071    &         46.872  &  $  49.012 \pm{0.945 } $          &    46.39   &    (  --, 5969  )   &    sy12 &  9.657     &     -0.401  &  --           \\
68   & SDSS143148.09+053558    &   2.097    &         47.092  &  $  46.011 \pm{1.013 } $          &    46.81   &    (  --, 1327  )   &    sy12 &  10.561    &     -0.885  &  --           \\
69   & SDSS142108.71+224117    &   2.185    &         47.07   &  $  58.384 \pm{1.033 } $          &    46.81  &    (  --, 5964  )   &    sy12 &  9.868     &     -0.188  &  --           \\
70   & SDSSJ152156.48+52023    &   2.208    &         47.609  &  $  9.1    \pm{0.6   } $          &    47.14   &    (  --, 5750  )   &    S15  &  9.999     &     0.007   &  --           \\
71   & UM645                   &   2.267    &         46.696  &  $  39.6   ^{+9.3   }_{-6.0   } $ &    46.31   &    (  --, 3966  )   &    S04  &  9.262     &     -0.085  &  761.41       \\
72   & SDSSJ170102.18+61230    &   2.287    &         46.632  &  $  18.7   ^{+3.8   }_{-3.2   } $ &    46.34   &    (  --, 5760  )   &    S04  &  9.601     &     -0.395  &  --           \\
73   & SDSSJ144245.66-02425    &   2.325    &         46.071  &  $  53.7   ^{+3.0   }_{-3.3   } $ &    46.03   &    (  --, 3661  )   &    N07  &  9.052     &     -0.156  &  --           \\
74   & SDSSJ115111.20+03404    &   2.337    &         45.359  &  $  47.2   ^{+2.3   }_{-2.1   } $ &    45.58   &    (  --, 5146  )   &    N07  &  9.123     &     -0.677  &  --           \\
75   & SDSSJ100710.70+04211    &   2.364    &         45.898  &  $  55     ^{+16.0  }_{-12.5  } $ &    45.17   &    (  --, 5516  )   &    N07  &  8.978     &     -0.942  &  --           \\
76   & UM642                   &   2.371    &         46.9    &  $  27.8   ^{+2.3   }_{-2.0   } $ &    46.29   &    (  --, 3925  )   &    S04  &  9.243     &     -0.086  &  --           \\
77   & SDSSJ125034.41-01051    &   2.398    &         45.895  &  $  72.3   \pm{0.2   } $          &    45.41   &    (  --, 5149  )   &    N07  &  9.038     &     -0.762  &  --           \\
78   & SDSSJ095141.33+01325    &   2.428    &         45.699  &  $  87.8   ^{+5.9   }_{-5.3   } $ &    45.55   &    (  --, 4297  )   &    N07  &  8.951     &     -0.535  &  --           \\
79   & SDSSJ101257.52+02593    &   2.433    &         46.094  &  $  34.9   ^{+0.6   }_{-0.1   } $ &    45.73   &    (  --, 3892  )   &    N07  &  8.955     &     -0.359  &  --           \\
80   & SDSS1138+0314           &   2.443    &         46.286  &  $  79.44  \pm{5.242 } $          &    44.81   &    (  --, 3930  )   &    A11  &  8.504     &     -0.828  &  --           \\
81   & UM629                   &   2.462    &         46.827  &  $  36     ^{+3.6   }_{-3.2   } $ &    46.56   &    (  --, 2621  )   &    S04  &  9.027     &     0.399   &  --           \\
82   & SDSSJ025438.37+00213    &   2.463    &         46.061  &  $  66.6   ^{+4.1   }_{-2.6   } $ &    45.85   &    (  --, 4164  )   &    N07  &  9.074     &     -0.358  &  --           \\
83   & SDSSJ024933.42-08345    &   2.491    &         46.655  &  $  51.4   \pm{0.2   } $          &    46.38   &    (  --, 5230  )   &    S04  &  9.537     &     -0.291  &  --           \\
84   & UM632                   &   2.499    &         46.851  &  $  34.431 \pm{3.449 } $          &    46.54   &    (  --, 3828  )   &    S04  &  9.346     &     0.060    &  1503.53      \\
85   & SDSSJ135445.66+00205    &   2.504    &         46.79   &  $  21.1   ^{+2.0   }_{-1.7   } $ &    46.49   &    (  --, 2627  )   &    S04  &  8.994     &     0.362   &  --           \\
86   & H1413+117               &   2.56     &         47.286  &  $  35.636 \pm{1.393 } $          &    45.63   &    (  --, 6700  )   &    A11  &  9.377     &     -0.881  &  --           \\
87   & Q0142-100               &   2.73     &         47.445  &  $  31.091 \pm{1.087 } $          &    46.27   &    (  --, 2700  )   &    A11  &  8.908     &     0.229   &  --           \\
88   & SDSSJ100428.43+00182    &   3.045    &         46.781  &  $  45.4   ^{+2.9   }_{-2.7   } $ &    46.44   &    (  --, 3442  )   &    S04  &  9.204     &     0.103   &  --           \\
89   & SBS1425+606             &   3.192    &         47.697  &  $  44.7   ^{+3.2   }_{-6.2   } $ &    47.38   &    (  --, 3144  )   &    S04  &  9.595     &     0.651   &  --           \\
90   & SDSSJ083700.82+35055    &   3.311    &         46.86   &  $  53.733 \pm{3.084 } $          &    46.62   &    (  6420, 8162)   &    J15  &  9.835     &     -0.349  &  --           \\
91   & SDSSJ210311.69-06005    &   3.336    &         46.473  &  $  91.644 \pm{5.953 } $          &    46.30    &    (  --, 6075  )   &    N07  &  9.627     &     -0.461  &  --           \\
92   & SDSSJ113838.26-02060    &   3.343    &         46.386  &  $  26.1   ^{+14.4  }_{-5.0   } $ &    45.79   &    (  --, 4562  )   &    N07  &  9.123     &     -0.467  &  --           \\
93   & SDSSJ210258.21+00202    &   3.345    &         46.125  &  $  42.6   ^{+7.9   }_{-6.4   } $ &    45.79   &    (  --, 7198  )   &    N07  &  9.519     &     -0.863  &  --           \\
94   & SDSSJ105511.99+02075    &   3.384    &         46.374  &  $  49.9   ^{+11.7  }_{-9.3   } $ &    45.70    &    (  --, 5424  )   &    N07  &  9.229     &     -0.662  &  --           \\
95   & SDSSJ083630.55+06204    &   3.4      &         46.294  &  $  14.9   ^{+29.7  }_{-6.4   } $ &    45.53   &    (  --, 3950  )   &    N07  &  8.868     &     -0.472  &  --           \\
96   & SDSSJ123743.08+63014    &   3.425    &         46.622  &  $  7.7    \pm{1.1   } $          &    46.35   &    (  --, 5200  )   &    S15  &  9.517     &     -0.301  &  --           \\
97   & SDSSJ173352.22+54003    &   3.425    &         47.418  &  $  22.1   ^{+16.0  }_{-9.6   } $ &    47.00      &    (  --, 3078  )   &    S04  &  9.387     &     0.480    &  13.9         \\
98   & SDSSJ115304.62+03595    &   3.432    &         46.531  &  $  12.8   ^{+6.7   }_{-3.6   } $ &    46.04   &    (  --, 5521  )   &    N07  &  9.414     &     -0.508  &  --           \\
99   & SDSSJ115935.64+04242    &   3.448    &         46.599  &  $  45.3   ^{+4.9   }_{-4.6   } $ &    45.92   &    (  --, 5557  )   &    N07  &  9.36      &     -0.573  &  --           \\
100  & SDSSJ153725.36-01465    &   3.452    &         46.484  &  $  34.5   ^{+1.5   }_{-1.4   } $ &    45.98   &    (  --, 3656  )   &    N07  &  9.026     &     -0.18   &  --           \\
101  & SDSSJ114153.34+02192    &   3.48     &         46.845  &  $  0.4    \pm{0.2   } $          &    46.55   &    (  --, 5900  )   &    S15  &  9.727     &     -0.31   &  11.8         \\
102  & SDSSJ164248.71+24030    &   3.48     &         46.933  &  $  31.321 \pm{3.644 } $          &    46.41   &    (  2500, 3001)   &    J15  &  8.911     &     0.365   &  --           \\
103  & SDSSJ150620.48+46064    &   3.504    &         46.893  &  $  25.047 \pm{2.621 } $          &    46.38   &    (  6980, 8919)   &    J15  &  9.788     &     -0.541  &  --           \\
104  & SDSSJ142243.02+44172    &   3.545    &         47.007  &  $  20.796 \pm{1.113 } $          &    47.18   &    (  6110, 7744)   &    J15  &  10.072    &     -0.026  &  --           \\
105  & SDSSJ120934.54+55374    &   3.573    &         47.13   &  $  7.802  \pm{1.401 } $          &    46.96   &    (  2500, 3001)   &    J15  &  9.186     &     0.64    &  --           \\
106  & SDSSJ075303.33+42313    &   3.59     &         47.125  &  $  30.149 \pm{1.053 } $          &    46.79   &    (  6240, 7919)   &    J15  &  9.895     &     -0.239  &  2645.33      \\
107  & SDSSJ120447.15+33093    &   3.616    &         46.37   &  $  126.312\pm{4.401 } $          &    46.97   &    (  7820,10062)   &    J15  &  10.181    &     -0.345  &  --           \\
108  & SDSSJ101336.37+56153    &   3.633    &         46.912  &  $  23.303 \pm{2.299 } $          &    46.99   &    (  6650, 8473)   &    J15  &  10.051    &     -0.194  &  --           \\
109  & SDSSJ144144.76+47200    &   3.633    &         46.779  &  $  63.594 \pm{10.716} $          &    46.56   &    (  2840, 3435)   &    J15  &  9.097     &     0.33    &  --           \\
110  & SDSSJ145408.95+51144    &   3.644    &         47.206  &  $  38.895 \pm{2.494 } $          &    47.08   &    (  4680, 5836)   &    J15  &  9.79      &     0.156   &  --           \\
111  & SDSSJ130348.94+00201    &   3.647    &         46.736  &  $  6.449  \pm{1.788 } $          &    46.73   &    (  6980, 8919)   &    J15  &  9.963     &     -0.366  &  --           \\
112  & SDSSJ015048.83+00412    &   3.702    &         46.885  &  $  27.598 \pm{6.089 } $          &    46.64   &    (  6470, 8229)   &    J15  &  9.852     &     -0.346  &  --           \\
113  & SDSSJ014049.18-08394    &   3.713    &         47.263  &  $  24.952 \pm{1.253 } $          &    46.96   &    (  5050, 6327)   &    J15  &  9.797     &     0.030    &  --           \\
114  & SDSSJ113307.63+52283    &   3.736    &         46.686  &  $  20.66  \pm{4.287 } $          &    46.64   &    (  2500, 3001)   &    J15  &  9.026     &     0.480    &  --           \\
115  & SDSSJ162520.31+22583    &   3.768    &         47.136  &  $  28.909 \pm{1.413 } $          &    46.66   &    (  5370, 6753)   &    J15  &  9.7       &     -0.174  &  --           \\
116  & SDSSJ012403.77+00443    &   3.834    &         47.123  &  $  39.35  \pm{2.64  } $          &    46.83   &    (  3760, 4627)   &    J15  &  9.475     &     0.221   &  --           \\
117  & SDSSJ144542.75+49024    &   3.875    &         47.288  &  $  56.39  \pm{3.353 } $          &    47.12   &    (  6570, 8364)   &    J15  &  10.105    &     -0.119  &  11.36        \\
118  & SDSSJ093554.45+52561    &   4.005    &         46.871  &  $  26.945 \pm{1.766 } $          &    46.78   &    (  3040, 3693)   &    J15  &  9.266     &     0.380    &  --           \\
119  & SDSSJ132420.83+42255    &   4.035    &         46.711  &  $  76.709 \pm{7.268 } $          &    46.65   &    (  3330, 4067)   &    J15  &  9.28      &     0.236   &  --           \\
120  & SDSSJ105756.28+45555    &   4.138    &         47.343  &  $  26.743 \pm{1.368 } $          &    47.24   &    (  4220, 5229)   &    J15  &  9.781     &     0.326   &  --           \\
121  & SDSSJ095511.32+59403    &   4.336    &         47.026  &  $  46.83  \pm{5.299 } $          &    46.81   &    (  3710, 4561)   &    J15  &  9.454     &     0.222   &  --           \\
122  & SDSSJ083946.22+51120    &   4.39     &         46.952  &  $  17.834 \pm{7.83  } $          &    46.71   &    (  6220, 7892)   &    J15  &  9.853     &     -0.276  &  285.1        \\
123  & SDSSJ010619.24+00482    &   4.449    &         47.002  &  $  59.795 \pm{5.143 } $          &    46.80    &    (  7750, 9967)   &    J15  &  10.089    &     -0.422  &  --           \\
124  & SDSSJ134743.29+49562    &   4.51     &         47.367  &  $  28.466 \pm{2.909 } $          &    46.97   &    (  6780, 8648)   &    J15  &  10.057    &     -0.221  &  --           \\
125  & SDSSJ163636.92+31571    &   4.559    &         47.029  &  $  12.52  \pm{5.536 } $          &    46.55   &    (  6660, 8486)   &    J15  &  9.832     &     -0.416  &  --           \\
126  & SDSSJ143835.95+43145    &   4.611    &         47.422  &  $  20.563 \pm{2.513 } $          &    47.14   &    (  4920, 6154)   &    J15  &  9.864     &     0.142   &  --           \\

\hline
\end{tabular}
\end{lrbox}
\scalebox{0.90}{\usebox{\tablebox}}
\end{table*}

\begin{table*}
\centering
\caption{Summary of the Spearman correlation coefficients and BCES Bisector fit results between \civ EW and \leddR for low-$z$ PG QSOs and all QSOs. Col.1: sample; Col.2: QSOs number; Col.3-4: the Spearman-rank correlation coefficient and the
probability of the null hypothesis; Col. 5-6: the intercept and the slope from BCES Bisector best-fitting relation, the value in brackets is error.}
\begin{tabular}{cccccc}
\hline
$Sample$  & $N$ & $r_s$  & $p$  & $a$ & $b$ \\
(1) &(2) &(3)& (4) & (5) &(6)\\
\hline
$low-$z$$     &$61$ & $-0.57$ & $2.71\times 10^{-8}$ &$1.28(0.08)$ & $-0.53(0.10)$\\
$all$  &$160$ & $-0.45$ & $1.52\times 10^{-11}$ &$1.35(0.05)$ & $-0.58(0.09)$\\
\hline
\end{tabular}
\label{table3}
\end{table*}

\end{document}